\documentclass[reprint,aps,showpacs,pra,nofootinbib]{revtex4-1}
  \usepackage{color}
  \usepackage{newlfont}
  \usepackage{graphicx}
  \usepackage{amssymb}
  \usepackage{amsmath}
  \usepackage{bm}
  \usepackage{verbatim}
  \usepackage[latin1]{inputenc}
  \usepackage{bbm}
  \usepackage{multirow}
  \usepackage[thinlines]{easytable}
  \usepackage[varg]{txfonts} 

\newcommand{\ket}[1]{\mbox{$ | #1 \rangle $}}
\newcommand{\bra}[1]{\mbox{$ \langle #1 | $}} 
\newcommand{\be}{\begin{equation}}
\newcommand{\ee}{\end{equation}}
\newcommand{\beq}{\begin{eqnarray}}
\newcommand{\eeq}{\end{eqnarray}}
\newcommand{\cm}{\text{cm}}

\DeclareMathOperator{\Tr}{Tr}

\begin{document}

\title{Quantifying continuous-variable realism}
\author{I. S. Freire}
\author{R. M. Angelo}
\affiliation{Department of Physics, Federal University of Paran\'a, P.O. Box 19044, 81531-980 Curitiba, Paran\'a, Brazil}

\begin{abstract} 
The debate instigated by the seminal works of Einstein, Podolsky, Rosen, and Bell put the notions of realism and nonlocality at the core of almost all philosophical and physical discussions underlying the foundations of quantum mechanics. However, while experimental criteria and quantifiers are by now well established for nonlocality, there is no clear quantitative measure for the degree of reality associated with continuous variables such as position and momentum. This work aims at filling this gap. Considering position and momentum as effectively discrete observables, we implement an operational notion of projective measurement and, from that, a criterion of reality for theses quantities. Then we introduce a quantifier for the degree of irreality of a discretized continuous variable which, when applied to the conjugated pair position-momentum, is shown to obey an uncertainty relation, meaning that quantum mechanics prevents classical realism for conjugated quantities. As an application of our formalism, we study the emergence of elements of reality in an instance where a Gaussian state is submitted to the dissipative dynamics implied by the Caldirola-Kanai Hamiltonian. In particular, at equilibrium, we make some links with the measurement problem and identify aspects that can be taken as the quantum counterpart for the notion of rest.
\end{abstract}

\maketitle

\section{Introduction}
\label{int}

Until the beginning of the twentieth century, classical physics nourished the belief that the physical properties of a system are all well defined at every instant in time, even when the system is ideally isolated from its surroundings. Quantum mechanics, however, tells us that this intuition about the physical reality actually emerges because we, observers, ``look'' at the systems through a vast number of projective measurements performed one after another almost instantaneously. This is what happens, for instance, when we collect the photons that are scattered by an object under observation. The resulting sequence of collapses keeps the values of physical quantities well defined and produces the preconception of an observer-independent physical reality. This feeling is further strengthened by the fact the macroscopic objects are barely disturbed by the measurement act and thus do not significantly deviate from their Newtonian trajectories. 

As has repeatedly been shown by experiments with isolated microscopic systems, such a classical notion of reality cannot be generally maintained. Perhaps the most emblematic experiment in this regard be the double-slit setup for matter (see, e.g.,~\cite{Tonomura1989,Arndt1999}), where massive particles are put in a coherent superposition of different locations and then produce an interference pattern. Here comes the conceptual difficulty: What can one say about the positional elements of reality of each particle in this experiment? Does a particle pass through both slits simultaneously, does it pass through as a wave, or is its position in a state of fundamental indefiniteness?

The implications of the superposition principle for the physical reality soon bothered the founding fathers of quantum theory. To emphasize the conundrum, Schr\"odinger showed that, governed by quantum laws, nature admits exotic scenarios where complex beings can be set in a sort of suspended reality which interpolates between states of ``being dead'' and ``being alive''~\cite{Schrodinger1935}. Approaching the issue from a different perspective, Einstein, Podolsky, and Rosen (EPR) put forward in 1935 a rationale defending that quantum mechanics could not be our ultimate description of nature~\cite{Einstein1935}. Taking locality as a cornerstone of physics and introducing a criterion of reality, they conceived superposition states for which, they claimed, incompatible observables would have simultaneous elements of reality, even though such elements cannot be predicted by quantum mechanics. Quantum mechanics was then regarded as incomplete. Ironically, Bell showed later that any theory aiming at completing quantum mechanics would be unavoidably nonlocal~\cite{Bell1964} (Bohmian mechanics~\cite{Bohm1952a,Bohm1952b} being a prominent illustration of this). In light of the substantial empirical evidence obtained from accurate loophole-free Bell tests~\cite{Hensen2015,Giustina2015,Shalm2015,Hensen2016,Rauch2018,Li2018}, it is fair to say nowadays that the fundamental premise of EPR, namely, locality, is unsustainable.

Discussions about the physical reality implied by the wave function were recently polarized in two main classes, both supported by a substantial amount of theoretical work. While on the one hand, $\psi$-ontic models aims at attaching to the wave function some realistic substance, on the other hand, $\psi$-epistemic models suggest that it actually represents the knowledge one has about an underlying reality. Specialized literature has by now cumulated a number of contributions in favor of both $\psi$-ontic~\cite{Pusey2012,Lewis2012,Colbeck2012,Hardy2013,Patra2013,Aaronson2013,Leifer2014,Barrett2014,Branciard2014,Ringbauer2015} and $\psi$-epistemic~\cite{Ballentine1970,Emerson2001,Spekkens2007,vanEnk2007,Harrigan2010,Bartlett2012,Spekkens2015} models, with a recent work arguing that quantum mechanics can be viewed as classical statistical mechanics with an ontic extension and an epistemic restriction~\cite{Budiyono2017}. Within the decoherence paradigm, where environmental models are provided to account for the disappearance of quantum superpositions, considerable progress has been made towards a deep understanding of the quantum measurement problem and the emergence of objective classicality in the framework of the quantum Darwinism~\cite{Zurek2009,Burke2010,Riedel2010,Horodecki2015,Brandao2015}. Also noteworthy is the framework according to which quantum physics is an elementary theory of information and, as such, some ontological status is to be given to the very notion of information~\cite{Zeilinger1999,Brukner1999,Brukner2003}.

Besides inciting a number of developments around the notions of entanglement and Bell nonlocality, EPR's criterion of reality also led to heated debates about the physical realism. Bohr's reply to EPR was given in terms of the complementarity principle~\cite{Bohr1935}, which defends that elements of reality of incompatible observables cannot be established in the same experiment, but only through mutually exclusive experimental arrangements. In Bohr's view, one cannot speak of the nature of microscopic systems before making a measurement, that is, physical reality only emerges upon interaction with a macroscopic apparatus. This perspective refutes EPR's rationale and elects the correlations generated in the experimental setup as an important mechanism responsible for the establishment of physical reality (see Ref.~\cite{Angelo2015} for a related discussion). Published in the same year as EPR's and Bohr's articles, Ruark's work~\cite{Ruark1935} pointed out that EPR's conclusion derived from a criterion that ``is directly opposed to the view held by many theoreticians, that a physical property of a given system has reality only when it is actually measured.'' Later, Redhead~\cite{Redhead1989} defended a subtle reformulation of EPR's criterion: ``If we can predict with certainty, or at any rate with probability one, the result of measuring a physical quantity at time $t$, then, at time $t$, there exists an element of reality corresponding to this physical quantity and having value equal to the predicted measurement result.'' This proposal aimed at softening the condition on the relativistic causality hypothesis. In 1996, Vaidman introduced a slightly different perspective on the issue. Realizing that a point common to many criteria of reality is the link with actual results of projective measurements, he proposed that ``for any definite result of a measurement there is a corresponding element of reality''~\cite{Vaidman1996}. Vaidman regarded as a ``definitive result'' the definite shift of the probability distribution of the pointer variable and thus suggested the following definition of elements of reality: ``If we are certain that a procedure for measuring a certain variable will lead to a definite shift of the unchanged probability distribution of the pointer, then there is an element of reality: the variable equal to this shift.'' With that, different shades of reality were attached to physical realism. A few years ago, Bilobran and Angelo (BA) introduced a measurement-based criterion of reality which allowed for the quantification of the degree of reality of a discrete-spectrum observable for a given multipartite quantum state~\cite{Bilobran2015}. This approach led to further foundational developments, as, for instance, the definition of the realism-based nonlocality~\cite{Gomes2018,Gomes2019}, which captures nonlocal aspects that are dramatically different from Bell nonlocality, and the derivation of an information-reality complementarity relation~\cite{Dieguez2018a}, which is shown to apply even to scenarios of weak disturbances. Interestingly, this framework received experimental corroboration via weak measurements implemented in a photonic platform~\cite{Mancino2018}.

With regard to the emergence of reality, a crucial aspect is recognized in Ref.~\cite{Dieguez2018a}, namely, that in all measurements, the degree of freedom that is intended to be measured is effectively discarded. In the Stern-Gerlach setting, for instance, the information about the spin is encoded in the spatial degree of freedom of the silver atom. After interacting with the magnet, the atom is set in the entangled state $\ket{+}\ket{\varphi_+}+\ket{-}\ket{\varphi_-}$ and then its position is registered (via some ionizing process) in a screen. The spin is then inferred from this position measurement. The discard of the spin subspace reduces the accessible state to $\ket{\varphi_+}\bra{\varphi_+}+\ket{\varphi_-}\bra{\varphi_-}$. If $\langle\varphi_+|\varphi_-\rangle=0$, then this state corresponds to a classical statistical mixture (with no interference terms in the position basis). In this case, the spin can be inferred with certainty, the measurement is ideally selective, and an element of reality emerges for the spin (even when nobody looks at the screen). In terms of the information-reality complementarity~\cite{Dieguez2018a}, the reality of the spin increases because information about it flows to the spatial degree of freedom. In fact, the narrative goes the other way around: because information about the atom position flows to the spin degree of freedom, which is inevitably discarded, an element of reality emerges for the position. This interpretation is consistent with the fact that no interference pattern is observed on the screen. In addition, it suggests, in agreement with the framework delineated in Ref.~\cite{Angelo2015}, that the wave-particle duality, widely accepted as a fundamental principle of quantum theory, to which both matter and radiation are submitted, can be phrased in terms of the dichotomy ``absence versus presence of quantum correlations.'' 

As illustrated in the aforementioned paradigmatic experiment, measurements of arbitrary physical quantities can always be reduced to a position measurement. Indeed, at the very last stage of any measurement, we always look at a ``pointer,'' that is, we invariably receive, from a physical carrier, information about the occurrence of a given event in a well-defined point in space-time. That is what happens, for instance, when a photon (or a sound wave) reaches an observer after being generated by a phosphorescent mark on a screen (or a click in a detector array) upon the arrival of a particle. Thus, within the context of the BA elements of reality, the question naturally arises whether one can quantify the extent to which the position of a system can be regarded as an element of physical reality. In spite of its acute foundational relevance, this problem has attracted little attention from the physical community, possibly because of the many technical and conceptual difficulties underlying this task. This work is devoted to solving this problem. Specifically, we want to extend the BA approach (which is briefly reviewed in Sec. II) to the domain of continuous variables and then look at the consequences implied by the derived measure for the elements of reality of canonically conjugated observables, such as position and momentum. As an application, we investigate the emergence of reality in the dissipative dynamics implied by the Caldirola-Kanai (CK) model, whose classical analog allows for the achievement of rest. Additionally, looking for a dynamical description of a position measurement, we expand the model in a way to allow for the analysis of an entanglement dynamics of a particle and a pointer.

\section{Basic concepts and uncertainty relation}

Recently, Bilobran and Angelo put forward a proposal for quantifying elements of reality of discrete-spectrum observables~\cite{Bilobran2015}. Here we present a brief review of the BA approach since it is the main platform for the present work. The whole idea is based on the single premise that upon the completion of a measurement of a given observable, say $A$, there exists an element of reality for $A$. The rationale goes as follows.

Let $\rho$ be an $n$-partite state on the Hilbert space $\mathcal{H}$, where $\mathcal{H=H_A\otimes H_B}$ with $\mathcal{H_B}=\bigotimes_{i=2}^n\mathcal{H}_i$ for $n\geqslant 2$ and $\mathcal{H=H_A}$ for $n=1$.
After measurement of an observable, $A=\sum_a aA_a$, with discrete spectrum $\{a\}$ and projectors $A_a=\ket{a}\bra{a}$ on $\mathcal{H_A}$, the state collapses to $\rho^{(a)}=(A_a\otimes\mathbbm{1}_{\cal B})\,\rho\,(A_a\otimes\mathbbm{1}_{\cal B})/p_a$, where $p_a=\Tr[(A_a\otimes\mathbbm{1}_{\cal B})\,\rho\,(A_a\otimes\mathbbm{1}_{\mathcal B})]$ and $\mathbbm{1}_\mathcal{B}=\bigotimes_{i=2}^n\mathbbm{1}_i$. According to the aforementioned premise, an element of reality then emerges for $A$. Mathematically, this is explicitly revealed by the projector appearing in $\rho^{(a)}=A_a\otimes\bra{a}\rho\ket{a}/p_a$. The adaptation of these formulas for the single-partite case ($n=1$) is straightforward (since in this case only one space is needed). Consider now that the outcome obtained by the experimentalist in each run of the experiment is not revealed to an external observer, who is asked to find out, via state tomography, what the post-measurement state is. Of course, the mere omission of the outcome can by no means alter the state of reality of $A$. The best description the ignorant observer can attain for the post-measurement ensemble through this procedure is
\be
\sum_ap_a\,\rho^{(a)}=\sum_a (A_a\otimes\mathbbm{1}_{\cal B})\,\rho\,(A_a\otimes\mathbbm{1}_{\mathcal B})=:\Phi_A(\rho).
\label{PhiA}
\ee 
$\Phi_A$ is a completely positive trace-preserving map written in terms of the operator-sum representation,  with Kraus operators $A_a$ satisfying $\sum_aA_a^{\dag}A_a=\mathbbm{1}_{\mathcal{A}}$. In this respect, $\Phi_A$ constitutes a particular form of quantum operation~\cite{Nielsen2000}. The unrevealed-measurement procedure thus leads to the state $\Phi_A(\rho)$, which, by virtue of the premise adopted, is to be interpreted as a state of reality for $A$. Note, in particular, that further unrevealed measurements of $A$ cannot change a state of reality $\rho'=\Phi_A(\rho)$, since $\Phi_A^2=\Phi_A$. This allows us to take 
\be
\Phi_A(\rho)=\rho\qquad\qquad \text{(BA criterion of realism)} 
\label{BAcriterion}
\ee 
as a formal statement of a scenario where $A$ is real for a given preparation $\rho$. With that, it is possible to employ the relative entropy $S(\rho||\sigma):=\Tr[\rho(\ln{\rho}-\ln{\sigma})]$ of states $\rho$ and $\sigma$ as an ``entropic metric'' to evaluate the so-called {\it irreality} $\mathfrak{I}(A|\rho)$ of $A$ given $\rho$, that is, the amount by which the BA realism is violated for the context defined by $A$ and $\rho$:
\be 
\mathfrak{I}(A|\rho):=S(\rho||\Phi_A(\rho))=S(\Phi_A(\rho))-S(\rho).
\label{frakI}
\ee 
Since $S(\rho||\sigma)$ is always nonnegative and equals 0 if and only if $\rho=\sigma$, then it is guaranteed that the irreality will vanish if and only if the BA realism occurs. Indeed, it is straightforward to check that $\mathfrak{I}(A|\Phi_A(\rho))=0$ for any $\rho$. Although one could have employed some norm to measure the distance between $\rho$ and $\Phi_A(\rho)$, the use of the entropic metric allows one to make insightful connections with some well-established quantities underlying the foundations of quantum information theory. In fact, as shown in Ref.~\cite{Bilobran2015}, irreality can be written in the form $\mathfrak{I}(A|\rho)=\mathfrak{I}(A|\rho_\mathcal{A})+D_A(\rho)$. The first parcel, $\mathfrak{I}(A|\rho_\mathcal{A})=S(\rho_\mathcal{A}||\Phi_A(\rho_\mathcal{A}))$, quantifies the amount of quantum coherence (with respect to the $A$ eigenbasis~\cite{Baumgratz2014}) that is encoded in the reduced state $\rho_\mathcal{A}=\Tr_\mathcal{B}\rho$. The second one, $D_A(\rho)=I_{\mathcal{A:B}}(\rho)-I_{\mathcal{A:B}}(\Phi_A(\rho))$, with $I_{\mathcal{A:B}}(\rho)=S(\rho||\rho_{\mathcal A}\otimes\rho_{\mathcal B})$ being the mutual information of $\rho$, is the basis-dependent quantum discord~\cite{Rulli2011}---a measure of correlations. In consonance with this remark, it can be checked directly from definition \eqref{frakI} that, for a single-partite state $\rho=\varrho_\mathcal{A}$ or a fully uncorrelated state $\rho=\varrho_\mathcal{A}\otimes\varrho_\mathcal{B}$, the irreality of an observable $A$ is entirely determined by the quantum coherence $\mathfrak{I}(A|\varrho_\mathcal{A})$.

While the irreality of $A$ vanishes for a state of reality $\Phi_A(\rho)$, this is not so if the preparation is a state of reality for an incompatible observable. That is, for $[A,A']\neq 0$ one has that $\mathfrak{I}(A|\Phi_{A'}(\rho))\geqslant 0$, with the equality holding only for $\rho$ being also a state of reality for $A$. This suggests that we cannot make the irrealities of incompatible observables vanish simultaneously. Interestingly, now we demonstrate that this is indeed the case. Using definition \eqref{frakI} along with the result $S(\rho)+S(\Phi_A\Phi_{A'}(\rho))\leqslant S(\Phi_A(\rho))+S(\Phi_{A'}(\rho))$, proven in Ref.~\cite{Dieguez2018a}, we can immediately demonstrate, for any $A$ and $A'$ acting on $\mathcal{H_A}$, that
\be
\mathfrak{I}(A|\rho)+\mathfrak{I}(A'|\rho)\geqslant I_{\mathcal{A|B}}(\rho),
\label{UR}
\ee 
where $I_{\mathcal{A|B}}(\rho)=\ln{d_{\mathcal{A}}}-S_{\mathcal{A|B}}(\rho)$, with $S_{\mathcal{A|B}}(\rho)=S(\rho)-S(\rho_{\mathcal{B}})$ and $\rho_{\mathcal{A}}=\Tr_{\mathcal{B}}(\rho)$. Since $S_{\mathcal{A|B}}$ is the conditional entropy, the lower bound $I_{\mathcal{A|B}}$ can be interpreted as the information available about the subsystem $\mathcal{A}$ from knowledge about the subsystem $\mathcal{B}$. This term can yet be written as
\be 
I_{\mathcal{A|B}}(\rho)=I(\rho_{\mathcal A}) +I_{\mathcal A:B}(\rho)=S\Big(\rho\big|\big|\tfrac{\mathbbm{1}}{d_{\mathcal{A}}}\otimes\rho_{\mathcal{B}}\Big),
\ee  
where $I(\rho_{\mathcal A})=\ln{d_{\mathcal{A}}}-S(\rho_{\mathcal{A}})$ is the information related to $\rho_{\mathcal{A}}$ and $\rho_{\mathcal{A(B)}}=\Tr_{\mathcal{B(A)}}(\rho)$. Inequality \eqref{UR} defines an uncertainty relation\footnote{This uncertainty relation can also be derived from some results in Ref. \cite{Rudnicki2018} as long as we require, in addition, the restriction that $A$ and $A'$ be {\it maximally incompatible observables}. This means that the eigenbases $\{\ket{a}\}$ and $\{\ket{a'}\}$ of $A$ and $A'$, respectively, must form mutually unbiased bases respecting $|\bra{a'}a\rangle|^2=d_{\mathcal A}^{-1}$.} for the irrealities of arbitrary observables $A$ and $A'$ on $\mathcal{H_A}$. The equality applies when $\rho=\Phi_{A'}\Phi_A(\rho)=\tfrac{\mathbbm{1}}{d_{\mathcal{A}}}\otimes\rho_{\mathfrak{B}}$ (a state of simultaneous reality for maximally incompatible observables $A$ and $A'$). This instance, however, provides no interesting insight, since all terms vanish. Consider, on the other hand, the pure state $\rho=\ket{\psi}\bra{\psi}$ of two equidimensional subsystems ($d_{\mathcal{A,B}}=d$), with Schmidt decomposition $\ket{\psi}=\sum_i\ket{i}\ket{i}/\text{\small $\sqrt{d}$}$. In this case $I(\rho_{\mathcal{A}})=0$ and $I_{\mathcal{A:B}}(\rho)=2S(\rho_{\mathcal{A}})=2\ln{d}$ (twice the entanglement of $\ket{\psi}$). This shows that, for pure states, entanglement prevents any two observables from having simultaneous reality.

The BA criterion for the occurrence of realism, \eqref{BAcriterion}, and the quantification of its violation, \eqref{frakI}, count by now with further developments and applications~\cite{Gomes2018,Gomes2019,Dieguez2018a,Mancino2018}. All these works are, however, strictly connected with discrete-spectrum observables. So far, this has been a mandatory specialization, as the BA approach fundamentally relies on projectors, whose definition is tricky for continuous variables. In what follows, we develop a scheme to suitably treat elements of reality associated with position and momentum.

\section{Discretization}

Here we show that one can directly apply the BA approach to the case of continuous-spectrum observables by making small adaptations. The main idea consists of keeping definitions \eqref{BAcriterion} and \eqref{frakI} intact and treating position and momentum as discrete quantities. This is, in fact, rather convenient for eventual comparisons with experimental contexts, wherein finite-resolution detectors operationally prevent the observation of genuine continuous behavior for whatever physical quantities. Let us then consider the eigenvalue relation $Q\ket{q_k}=q_k\ket{q_k}$ for the operator $Q$ (describing a discrete generalized coordinate), where $\ket{q_k}$ is the nonnormalized eigenvector associated with the eigenvalue $q_k=k\,\delta q$, with $k$ being an integer and $\delta q$ an (operational) resolution of the position space. Within this model, $\delta q>0\in\mathbbm{R}$ is a (small) free parameter with dimensional unit of length. We then introduce
\be 
\langle q_{k}\ket{q_{k'}}=\frac{\delta_{kk'}}{\delta q}
\label{<q|q>}
\ee 
for the scalar product of the space, where $\delta_{kk'}$ is the Kronecker delta function. Clearly, this relation has the correct dimensional unit and leads to the Dirac delta function in the continuous variable limit $(\delta q\to 0)$. As projectors we propose
\be 
\Pi_k=\delta q\,\ket{q_k}\bra{q_k}
\label{projector_q}
\ee 
satisfying
\be 
\Pi_k\Pi_{k'}=\delta_{kk'}\Pi_k\qquad\quad \text{and} \qquad\quad \sum_{k=-L_q}^{L_q}\Pi_k=\mathbbm{1}.
\label{1_q}
\ee 
The parameter $L_q$, to be posteriorly fixed as an integer function of $\delta q$, defines the dimension $2L_q+1$ of the discretized space. Its introduction will prove relevant for the consistency of the method. The completeness relation given above allows one to expand any vector $\ket{\psi}$ as
\be 
\ket{\psi}=\sum_{k=-L_q}^{L_q}\delta q\,\,\psi(q_k)\,\ket{q_k},
\ee 
with amplitude $\psi(q_k)=\langle q_k\ket{\psi}$ and probability $|\psi(q_k)|^2\,\delta q$. An analog discretization scheme can be postulated for the momentum space. To this end, we introduce
\be 
\langle p_{l}\ket{p_{l'}}=\frac{\delta_{ll'}}{\delta p} \qquad\quad \text{and}\qquad\quad \Xi_l=\delta p\,\ket{p_l}\bra{p_l},
\label{projector_p}
\ee 
such that 
\be 
\Xi_l\Xi_{l'}=\delta_{ll'}\Xi_l \qquad\quad \text{and}\qquad\quad \sum_{l=-L_p}^{L_p}\Xi_l=\mathbbm{1}.
\label{1_p}
\ee 
$\delta p$ and $L_p$ play the roles of momentum resolution and momentum space dimension, respectively. Using the $\{\ket{p_l}\}$ representation we can expand any $\ket{\psi}$ as
\be 
\ket{\psi}=\sum_{l=-L_p}^{L_p}\delta p\,\,\psi(p_l)\,\ket{p_l},
\ee 
with amplitude $\psi(p_l)$ and probability $|\psi(p_l)|^2\delta p$.

In order to obtain interrelations for the various free parameters of the model, we require the validity of the usual algebra~\cite{Sakurai2010} associated with the canonical couple $\{Q,P\}$. Let $\mathcal{T}(\delta):=\mathbbm{1}-i P\delta/\hbar$ be the standard translation operator associated with an infinitesimal displacement $\delta\geqslant \delta q$ and generator $P$. By demanding that $\mathcal{T}(\delta q)\ket{q_k}=\ket{q_k+\delta q}=\ket{q_{k+1}}$ we directly find $[Q,\mathcal{T}(\delta q)]\ket{q_k}=\delta q\ket{q_k}$. Using the explicit form of $\mathcal{T}(\delta q)$ we obtain $[Q,P]=i\hbar$, which confirms that the translation generator $P$ is indeed the momentum operator satisfying $P\ket{p_l}=p_l\ket{p_l}$, with $p_l=l\,\delta p$. Since $P$ is Hermitian, all the properties expected for $\mathcal{T}$ are (up to the first-order approximation with respect to  $\delta$) validated; in particular, one shows that $\mathcal{T}(-\delta)=\mathcal{T}^{\dag}(\delta)$ and $\mathcal{T}(\delta)\mathcal{T}(\delta')=\mathcal{T}(\delta+\delta')$. Also, from $\bra{q_k}\mathcal{T}(\delta q)\ket{\psi}=\bra{q_k+\delta q}\psi\rangle=\psi(q_k)+\delta q\,\psi'(q_k)$ we have
\be 
\bra{q_k}P\ket{\psi}=\frac{\hbar}{i}\,\psi'(q_k)=\frac{\hbar}{i}\left[\frac{\psi(q_k+\delta q)-\psi(q_k)}{\delta q}\right],
\label{derivative}
\ee 
which identifies momentum in the position representation with a discrete derivative with respect to  position, as expected. In the continuous regime $(\delta q\to 0)$, specializing $\ket{\psi}=\ket{p}$ above gives, as the solution, the plane wave $\langle q|p\rangle=e^{i q p/\hbar}/\sqrt{2\pi\hbar}$. In the discrete regime, we still have (up to  first order in $\delta q$)
\be 
\langle q_k|p_l\rangle=\frac{e^{i\,q_k p_l/\hbar}}{\sqrt{2\pi\hbar}}=\frac{e^{i\,k\,l\,\delta q\,\delta p/\hbar}}{\sqrt{2\pi\hbar}}.
\ee 
Now, using this relation along with \eqref{<q|q>} and \eqref{1_p} yields
\be 
\delta q\,\langle q_k\ket{q_{k'}}=\frac{\delta q\,\delta p}{2\pi\hbar}\sum_{l=-L_p}^{L_p} e^{i\,l (k-k') \delta q\,\delta p/\hbar} =\delta_{kk'}.
\label{scalar}
\ee 
This formula can be directly compared with the identity
\be 
\frac{1}{\xi}\sum_{l=-\frac{\xi-1}{2}}^{\frac{\xi-1}{2}} e^{i\,2\pi\,l\,(k-k')/\xi}=\delta_{kk'}
\label{deltakk'},
\ee 
where $\xi$ is an odd integer and $\{k,k'\}\in\left[-\frac{\xi-1}{2},\frac{\xi-1}{2}\right]$, to produce $L_p=(\xi-1)/2$ and 
\be 
\xi=\frac{2\pi\hbar}{\delta q\,\delta p}.
\label{xi}
\ee 
By virtue of the condition required for $\xi$, this result constraints the values admissible for the resolutions $\delta q$ and $\delta p$. With an analogous procedure, we use relations \eqref{1_q} and \eqref{projector_p} to obtain $L_q=(\xi-1)/2=L_p=:L$. From the above, we have $\xi=2L+1$, which shows that the dimension of the discretized spaces is determined by the resolutions. Moreover, in order for one to have $L>0$, it is necessary that $\xi >1$, which implies that $\delta q\,\delta p<2\pi\hbar$. This adds an important limitation to the resolutions allowed in our discrete model. Hereafter, the limits of the summations, whenever omitted, are to be taken as $\pm L$. 

\subsection{Application for a Gaussian state}

The importance of Gaussian states for physics does not need emphasis. In the present work, theses states also play a distinctive role as shown later, especially because of their analytical properties. It is, therefore, instructive to discuss the implications and limitations of the discretization method in this context. The standard continuous-variable minimum-uncertainty Gaussian state can be written as 
\be
\ket{\psi}=\int_{-\infty}^{\infty} dq\,\psi(q-\bar{q})\,\ket{q}, \qquad
\psi(q)=\frac{e^{-\frac{q^2}{4(\Delta q)^2}}e^{i\bar{p}q/\hbar}}{[2\pi(\Delta q)^2]^{1/4}},
\label{CGaussian}
\ee
for which one shows that
\be
\begin{array}{lll}
\langle Q\rangle=\bar{q}, & \qquad & \Delta Q=\sqrt{\langle Q^2\rangle-\langle Q\rangle^2}=\Delta q, \\ \\
\langle P\rangle=\bar{p}, & & \Delta P=\sqrt{\langle P^2\rangle-\langle P\rangle^2}=\frac{\hbar}{2\Delta q}.
\end{array}
\ee
To migrate to the discrete model, we introduce 
\be
\ket{\psi}=\sum_{k}\sqrt{\delta q}\,\,\psi_{k-\bar{k}}\,\ket{k},\qquad \psi_k=\frac{e^{-\frac{k^2}{4\Delta_q^2}}e^{i 2\pi\,\bar{l}k/\xi}}{\sqrt{N}},
\label{DGaussian}
\ee
with $\ket{k}\equiv\ket{q_k}$, $\Delta_q\equiv \Delta q/\delta q$, $\bar{q}=\bar{k}\,\delta q$, and $\bar{p}=\bar{l}\,\delta p$. We opted to deal with a dimensionless amplitude $\psi_k$, which explains the appearance of the term $\sqrt{\delta q}$ instead of $\delta q$. The normalization term, which is fixed through $\langle \psi|\psi\rangle=1$, can be accurately approximated as
\be 
N=\sum_k |\psi_{k-\bar{k}}|^2\cong \sum_{k=-\infty}^{\infty}e^{-\frac{k^2}{2\Delta_q^2}}=\vartheta_3\Big(0,e^{-\frac{1}{2\Delta_q^2}}\Big)=: N_{\Delta_q},
\label{N}
\ee 
where $\vartheta_3(z,q)=\sum_{k=-\infty}^{\infty}q^{k^2}e^{i 2kz}$, with $z\in\mathbb{C}$, stands for the Jacobi theta function. Further analyses allow one to obtain the analytical approximation
\be
N_{\Delta_q}\cong \left\{ 
\begin{array}{lll} 1+2\left(e^{-\frac{1}{2\Delta_q^2}}+e^{-\frac{4}{2\Delta_q^2}}+e^{-\frac{9}{2\Delta_q^2}} \right) & & \text{if}\,\,0\!\leqslant\! \Delta_q\! <\! 1, \\ \\ \sqrt{2\pi\Delta_q^2} && \text{if}\,\,\Delta_q\!>\!1,
\end{array}\right.
\label{NDq}
\ee
which never implies an error greater than $0.0271\%$ with respect to $\vartheta_3\big(0,\exp(-1/2\Delta_q^2)\big)$ for all $\Delta_q\geqslant 0$. The quality of these results increases as the continuous limit $\delta q\,\delta p\to 0$ $(\xi\to \infty)$ is approached. 

\begin{figure}[htb] 
\centerline{\includegraphics[scale=0.45]{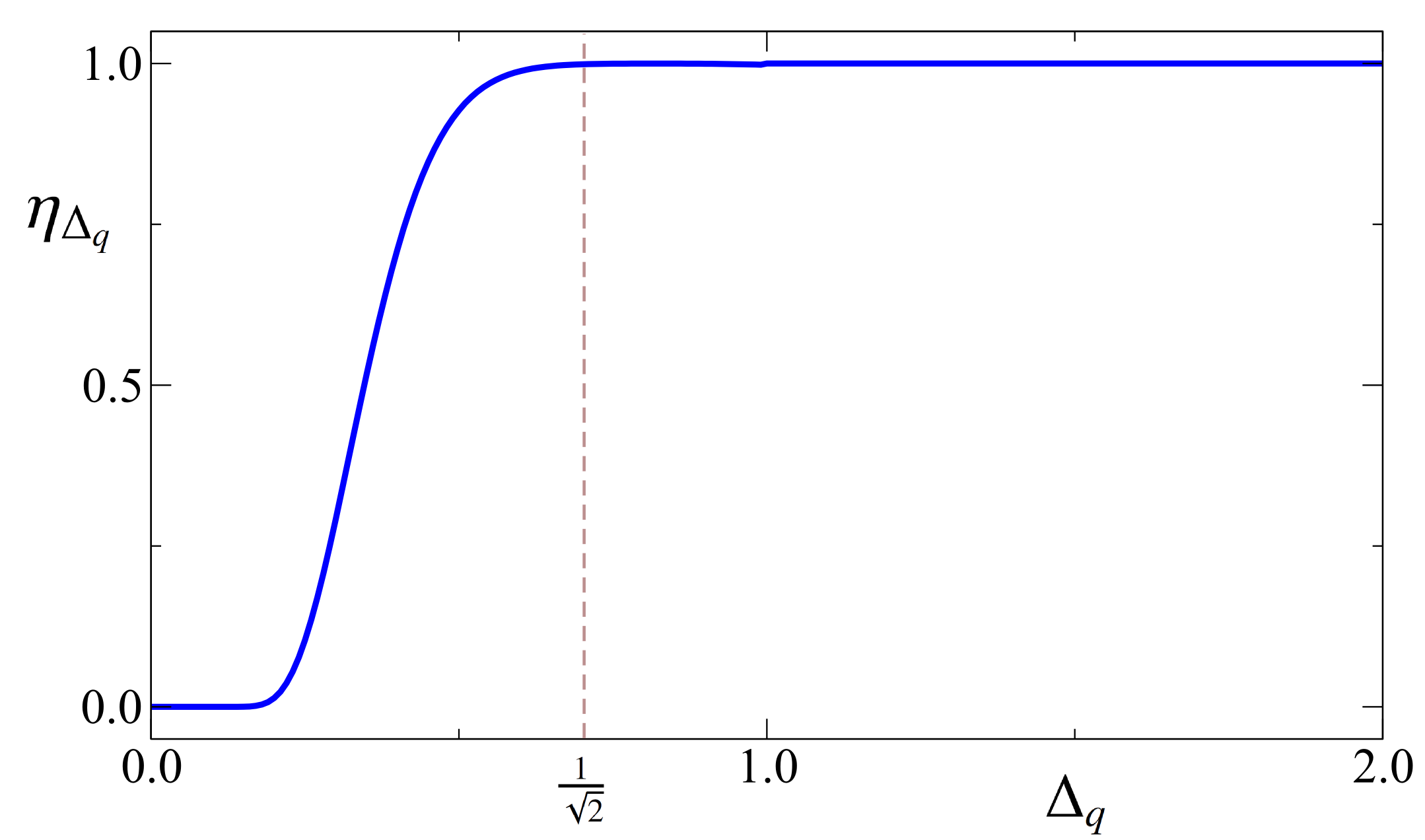}}
\caption{The representative $\eta_{\Delta_q}$ (solid blue line) of our discretized model for the uncertainty relation $2\,\Delta Q\,\Delta P/\hbar$ associated with a minimum uncertainty state as a function of the dimensionless width $\Delta_q=\Delta q/\delta q$. For $\Delta_q\geqslant 1$, one finds that $\eta_{\Delta_q}=1$, in full agreement with the continuous description. For $\Delta_q=1/\sqrt{2}$ (represented by the dashed brown line), the violation of the uncertainty principle is still negligible, since $\eta_{\Delta_q}\cong 0.9989$.}
\label{fig1}
\end{figure}

Having calculated the normalization, we can use the discretized wave function, \eqref{DGaussian}, to assess the quality of the resulting statistics. Employing the above approximations and the discrete derivative, \eqref{derivative}, and preserving only leading terms with respect to $\delta q$, one can check that
\beq
\langle Q \rangle &=& \sum_k\psi_{k-\bar{k}}^*\,\,(k\,\delta q)\,\,\psi_{k-\bar{k}}\cong \delta q\sum_{k=-\infty}^{\infty}\psi_{k}^*\,\,(k+\bar{k})\,\,\psi_{k}=\bar{k}\,\delta q, \nonumber \\
\langle P \rangle &=& \sum_k\psi_{k-\bar{k}}^*\,(-i\hbar)\,\psi_{k-\bar{k}}'\cong\bar{l}\,\delta p+i\left[\frac{\hbar}{8\Delta q}\left(\frac{\delta q}{\Delta q}\right)+\bar{l}\,\delta p\frac{\delta q\delta p}{2\hbar}\right],\nonumber \\ 
\langle Q^2\rangle &=& \sum_k \psi_{k-\bar{k}}^* \left(k\,\delta q\right)^2 \psi_{k-\bar{k}}\cong (\bar{k}\,\delta q)^2+(\delta q)^2\sum_{k=-\infty}^{\infty} k^2 |\psi_k|^ 2,\nonumber \\
\langle P^2\rangle &=& \sum_k\psi_{k-\bar{k}}(-i\hbar)^2\psi_{k-\bar{k}}''=(\bar{l}\,\delta p)^2+\left(\frac{\hbar}{2\Delta q}\right)^2,\nonumber
\eeq
which can always be identified to their continuous-case counterparts within certain approximations. Taking the identity $k^2\exp[-k^2/(2\Delta_q^2)]=-2\Delta_q^2\lim_{\lambda\to 1}\frac{d}{d\lambda}\,\exp[-\lambda k^2/(2\Delta_q^2)]$ it is possible to advance the computation for $\langle Q^2\rangle$ and for the uncertainties $\Delta Q$ and $\Delta P$, whose product can be written as
\be 
\left(\frac{\Delta Q\,\Delta P}{\hbar/2}\right)^2=\sum_{k=-\infty}^{\infty}\frac{k^2|\psi_k|^2}{\Delta_q^2} =-\frac{2}{N_{\Delta_q}}\lim_{\lambda\to 1}\frac{d}{d\lambda}N_{\frac{\Delta_q}{\sqrt{\lambda}}}=:\eta^2_{\Delta_q}.
\label{URqp}
\ee 
With approximation \eqref{NDq}, the function $\eta_{\Delta_q}$ can be analytically computed. Its behavior is presented in Fig.~\ref{fig1} as a function of $\Delta_q=\Delta q/\delta q$, showing that the discretized model is in full agreement with the continuous one as long as the state width $\Delta q$ is not appreciably smaller than the resolution $\delta q$.
%

\section{Quantifying realism violations} 

Having developed the discretized model for position and momentum, we can now apply the BA formalism to quantify the degree of irreality of these quantities for a given preparation $\rho$. The unrevealed-measurement map \eqref{PhiA} is now written as $\Phi_Q(\rho)=\sum_k\Pi_k\rho\Pi_k$, with projectors $\Pi_k$ defined by Eq. \eqref{projector_q} and $k$ running from $-L$ to $L=(\xi-1)/2$. The construction of $\Phi_P(\rho)$ is made in complete analogy by use of the projectors \eqref{projector_p}. It can be checked that $\Phi_R(\rho)$, with $R\in \{Q,P\}$, preserves all the desired properties; in particular; it is a completely positive trace-preserving map and $\Phi_R^2=\Phi_R$. The BA criterion of realism, \eqref{BAcriterion}, expresses itself for position and momentum in the form $\Phi_R(\rho)=\rho$. It then follows that the degree of irreality of $R$ can be diagnosed as
\be 
\mathfrak{I}(R|\rho)=S(\Phi_R(\rho))-S(\rho)\qquad (R\in\{Q,P\}).
\label{frakI_R}
\ee 
The form of the von Neumann entropy, $S(\rho)=-\Tr(\rho\ln{\rho})$, is preserved in the discretized model provided we write the trace operation as $\Tr(O)=\sum_k\delta q\bra{q_k}O\ket{q_k}=\sum_l\delta p\bra{p_l}O\ket{p_l}$, for a generic operator $O$. In addition, one shows that $S=0$ and $S=\ln{D}$ (with $D=2L+1$) for pure and maximally mixed states, respectively, in agreement with the continuous-variable formulation. 

\subsection{Examples}

It is instructive to compute the irreality for some simple cases. Consider as the first example a uniform state given by
\be 
\ket{\psi}=\sum_{k=-\bar{\kappa}}^{\bar{\kappa}}\sqrt{\delta q}\,\,\psi_k\,\ket{k},\qquad \psi_k=\frac{1}{\sqrt{\Delta_q}},
\ee 
where $\bar{\kappa}=(\Delta_q-1)/2$ and $\Delta_q=2n+1$ with $n\in\mathbb{N}$. It follows that
\be 
\Phi_Q(\rho)=\sum_k\Pi_k\,\ket{\psi}\bra{\psi}\,\Pi_k=\sum_k|\psi_k|^2\Pi_k=\frac{\mathbbm{1}}{\Delta_q}.
\ee 
Since $\Phi_Q(\rho)$ is a statistical mixture with eigenvalues $1/\Delta_q$ and $S(\ket{\psi})=0$, we obtain 
\be 
\mathfrak{I}(Q|\rho)=\ln{\Delta_q} \qquad\qquad \text{(uniform state)}.
\label{IrrUniform}
\ee 
This shows that the irreality increases with the width of the superposition, in this case also being a direct measure of quantum coherence \cite{Bilobran2015}.

For our second example, we return to the Gaussian state, \eqref{DGaussian}. In this case, we find 
\be 
\Phi_Q(\rho)=\sum_k|\psi_k|^2\Pi_k\cong \frac{1}{N_{\Delta_q}}\sum_{k=-\infty}^{\infty}e^{-\frac{k^2}{2\Delta_q^2}}\Pi_k,
\ee 
whose eigenvalues read $\exp\left[-k^2/(2\Delta_q^2)\right]/N_{\Delta_q}$. This leads to the irreality
\be 
\mathfrak{I}(Q|\rho)=-\sum_{k=-\infty}^{\infty}\frac{e^{-\frac{k^2}{2\Delta_q^2}}}{N_{\Delta_q}}\ln{\left(\frac{e^{-\frac{k^2}{2\Delta_q^2}}}{N_{\Delta_q}}\right)}=\ln{N_{\Delta_q}}+\frac{\eta^2_{\Delta_q}}{2},
\ee
where relations \eqref{N} and \eqref{URqp} have been used. With relations \eqref{NDq} and \eqref{URqp} derived for $N_{\Delta_q}$ and $\eta_{\Delta_q}$, respectively, one verifies that $\mathfrak{I}(Q|\rho)\to 0$ as $\Delta_q\to 0$, always preserving the positivity of the irreality. However, as pointed out above, in order not to violate the uncertainty principle we have to confine ourselves to $\Delta_q\geqslant 1$, the domain in which $\eta_{\Delta_q}=1$ and $N_{\Delta_q}=\sqrt{2\pi}\,\Delta_q$. With that, we finally obtain
\be 
\mathfrak{I}(Q|\rho)=\ln\left(\sqrt{2\pi e}\,\Delta_q \right) \qquad\qquad \text{(Gaussian state)}.
\label{IrrGaussian}
\ee 
It is interesting to note that the probability of finding the particle in the range $\left(\langle Q\rangle-\tfrac{d}{2},\langle Q\rangle+\tfrac{d}{2}\right)$ for a Gaussian state with root mean square $\Delta q$ results in 0.383 for $d=\Delta q$ and 0.961 for $d=\sqrt{2\pi e}\,\Delta q$. In this sense, $\sqrt{2\pi e}\,\Delta q$ can be viewed as a better candidate for discriminating the ``effective width'' of the wave-packet with respect to the resolution $\delta q$. This observation unifies results \eqref{IrrUniform} and \eqref{IrrGaussian}.

\subsection{Position-momentum uncertainty relation}

Since the wave-packet considered above manifests itself as a Gaussian distribution also in the momentum representation, it is clear that we should have $\mathfrak{I}(P|\rho)=\ln{\big(\sqrt{2\pi e}\,\Delta_p \big)}$. It follows that $\mathfrak{I}(Q|\rho)+\mathfrak{I}(P|\rho)=\ln{\big(2\pi e\,\Delta_q\Delta_p \big)}$, which is in agreement with results reported in conceptually different contexts.\footnote{Since $S(\rho)=0$ and $\Delta q\Delta p=\hbar/2$ for the state under consideration, the present result can be written in the form $H_q +H_p=\ln{\big(\tfrac{\pi e\hbar}{\delta q \delta p}\big)}$, which was found in Ref. \cite{Birula2006}, where $H_q=S(\Phi_Q(\rho))$ and $H_p=S(\Phi_P(\rho))$ are the Shannon entropies associated with probability distributions for the classical variables $q$ and $p$. Consistency with  result \eqref{UR} is checked by writing $\mathfrak{I}(Q|\rho)+\mathfrak{I}(P|\rho)=\ln{(e\xi/2)}>\ln{\xi}=I(\rho)$. Recall that there are no correlations in this case and that $\xi$ accounts for the dimension of the space.} As previously mentioned, in order to keep the discretized model consistent with Heisenberg's uncertainty, we should demand that $\Delta_q\geqslant 1$ and $\Delta_p\geqslant 1$, which implies that 
\be 
\mathfrak{I}(Q|\rho)+\mathfrak{I}(P|\rho)\geqslant \ln{\left(2\pi e\right)}.
\label{URQP} 
\ee 
Once this lower bound has been derived for the minimum-uncertainty state, we expect that such inequality will be valid in general. This tells us that we can never prepare a pure state $\rho$ for which position and momentum are simultaneous elements of reality. We see, therefore, that Heisenberg's uncertainty relation imposes severe restrictions to the classical notion of realism, which here is written as $\mathfrak{I}(Q|\rho)=\mathfrak{I}(P|\rho)=0$.

\section{Quantum-mechanical rest}

One of the tenets of quantum measurement theory prescribes that the realization of sequential measurements of a given observable must always produce the same outcome revealed by the first of these measurements, as long as the system is not allowed to dynamically evolve between two contiguous measurements. If the observable under consideration is the position of a quantum particle, then we might conclude that such a sequential protocol would be able to confine the particle to some state of rest. Of course, due to Heisenberg's uncertainty principle, which implies full indefiniteness for the canonical momentum after each position measurement, the resulting state might not be strictly compatible with our classical notion of rest. However, as far as position and velocity are the figures of merit, it is still possible to find classical rest---with emergent elements of reality for both quantities simultaneously---even departing from a strictly quantum substratum. The aim of this section is to make this point through a perspective according to which rest can be achieved through an overdamped quantum dynamics. 

Our argument is constructed by use of the CK model~\cite{Caldirola1941,Kanai1948}, which {\it effectively} implements the dissipative dynamics of a block of mass $m$ attached to a spring with elastic constant $k$. The classical time-dependent Hamiltonian reads
\be 
h_t=\frac{p^2}{2m} e^{-2\tau}+\frac{kq^2}{2} e^{2\tau},
\label{ht}
\ee 
where $\tau=\lambda t$ is a dimensionless time, $t$ is the physical time, and $\lambda$ is a frequency that determines the dissipation rate. Hamiltons's equations of motion lead to $\ddot{q}+2\lambda\dot{q}+\omega^2q=0$, with $\omega^2=k/m$, where one can recognize a velocity-dependent term typical of damped motion. Direct integration of the equations of motion, with initial conditions $q_0$ and $p_0$, yield
\begin{subequations}
\beq  
q_t&=& q_0\,e^{-\tau}\left[\cosh{\left(\zeta \tau \right)}+\left(1+\frac{p_0^2/m}{\varepsilon_0}\right)\frac{\sinh{(\zeta\tau)}}{\zeta}\right], \\
p_t&=&p_0\,e^{\tau}\left[\cosh{\left(\zeta \tau \right)}-\left(1+\frac{k q_0^2}{\varepsilon_0}\right)\frac{\sinh{(\zeta\tau)}}{\zeta}\right],
\eeq
\end{subequations}  
where $\zeta=(1-\omega^2/\lambda^2)^{1/2}$, and $\varepsilon_0=\lambda q_0p_0$. Despite the notably divergent form of the canonical momentum, we may identify scenarios typical of mechanical rest if we look at the velocity $v_t=dq_t/dt$. In Table \ref{tab1}, the asymptotic behaviors ($t\to\infty$) of some physical quantities are presented as a function of the dimensionless time $\tau$ for distinct regimes of damping. It is clear that the block will eventually come to a fixed position with no velocity and no kinetic energy. The potential energy accumulated by the spring, $V_t=kq_t^2/2$, will be fully suppressed as well. It is then evident that $h_t$, which may increase with time, is not to be taken as the energy of the oscillator; this is the case only if $\lambda=0$. Instead, it may be interpreted as the total energy of a system composed of the oscillator (block $+$ spring) and an environment that drains the mechanical energy of the oscillator while receiving some energy supply from an external source\footnote{Tricky to interpret, the Caldirola-Kanai model is sometimes regarded as referring to an oscillator of increasing mass, which is justified by the term $me^{2\tau}$ appearing in the Hamiltonian, \eqref{ht}. On the other hand, the term $ke^{2\tau}$ could be viewed as describing a trapping potential of increasing strength, which demands some energy supply. In any case, the net effect is of an effective dissipative dynamics, as indicated by the equation of motion $\ddot{q}+2\lambda\dot{q}+\omega^2q=0$.}. Also noteworthy is the dramatic difference between velocity and canonical momentum. Indeed, from Hamilton's equation we have $mv=p\,e^{-2\tau}$.
\begin{table}[htb]
  \caption{Asymptotic behaviors ($t\to\infty$) of physical quantities as a function of the dimensionless time $\tau=\lambda t$ in three distinct regimes of damping for the CK model. The asymptotic behaviors for the kinetic and potential energies directly follow from $K_t\propto v_t^2$ and $V_t\propto q_t^2$, respectively. In all cases, mechanical rest, as defined by Eq. \eqref{Crest}, is achieved.}
  \label{tab1}
  \begin{center}
    \begin{tabular}{lccccccc}\hline\hline
        \text{\bf Regime}  & \qquad & $q_t$ & $p_t$ & $v_t$ & $h_t$  \\ 
      \hline
      \text{Underdamped} $(\lambda <\omega)$ & & $e^{-\tau}$ & $e^{\tau}$ & $e^{-\tau}$ & \text{oscillates} \\
      \text{Critically damped} $(\lambda=\omega)$ & & $\tau\,e^{-\tau}$ & $\tau\,e^{\tau}$ & $\tau\,e^{-\tau}$ & $\tau^2$ \\
      \text{Overdamped} $(\lambda>\omega)$  & & $e^{-(1-\zeta)\tau}$ & $e^{(1+\zeta)\tau}$ & $e^{-(1-\zeta)\tau}$ & $e^{2\zeta \tau}$ \\   
      \hline\hline
    \end{tabular}
  \end{center}
\end{table}
It is immediately seen that mechanical rest can be generally claimed to occur for the block in the CK model since
\be 
\lim_{t\to\infty} \left(q_t,v_t \right)=\boldsymbol{0}.
\label{Crest}
\ee 

The direct quantization of the classical model gives
\be 
H_t=\frac{P^2}{2m}e^{-2\tau}+\frac{kQ^2}{2}e^{2\tau},
\label{QCKH}
\ee 
with $[Q,P]=i\hbar$. Using Heisenberg's picture we can show that $\ddot{Q}_H+2\lambda\dot{Q}_H+\omega^2Q_H=0$, with $\omega^2=k/m$, where $Q_H=U_t^{\dag}QU_t$ and $U_t=\exp\left[-\tfrac{i}{\hbar} \int_0^t dt'H_{t'}\right]$. This shows that, for a well-localized initial state $\rho_0$ and $\lambda$ sufficiently large, the mean value $\langle Q\rangle_t=\Tr[\rho_0Q_H]$ evolves in time as a typical trajectory of a damped motion, so that the previously studied classical behavior will approximately apply (Ehrenfest's theorem). However, since we are interested in analyzing whether and how the elements of reality emerge from the quantum dynamics, the study of the centroid does not suffice. In particular, because the irreality quantifier, \eqref{frakI}, is a state variable, the Schr\"odinger picture should be preferred.

\begin{figure}[htb]
\centerline{\includegraphics[scale=0.13]{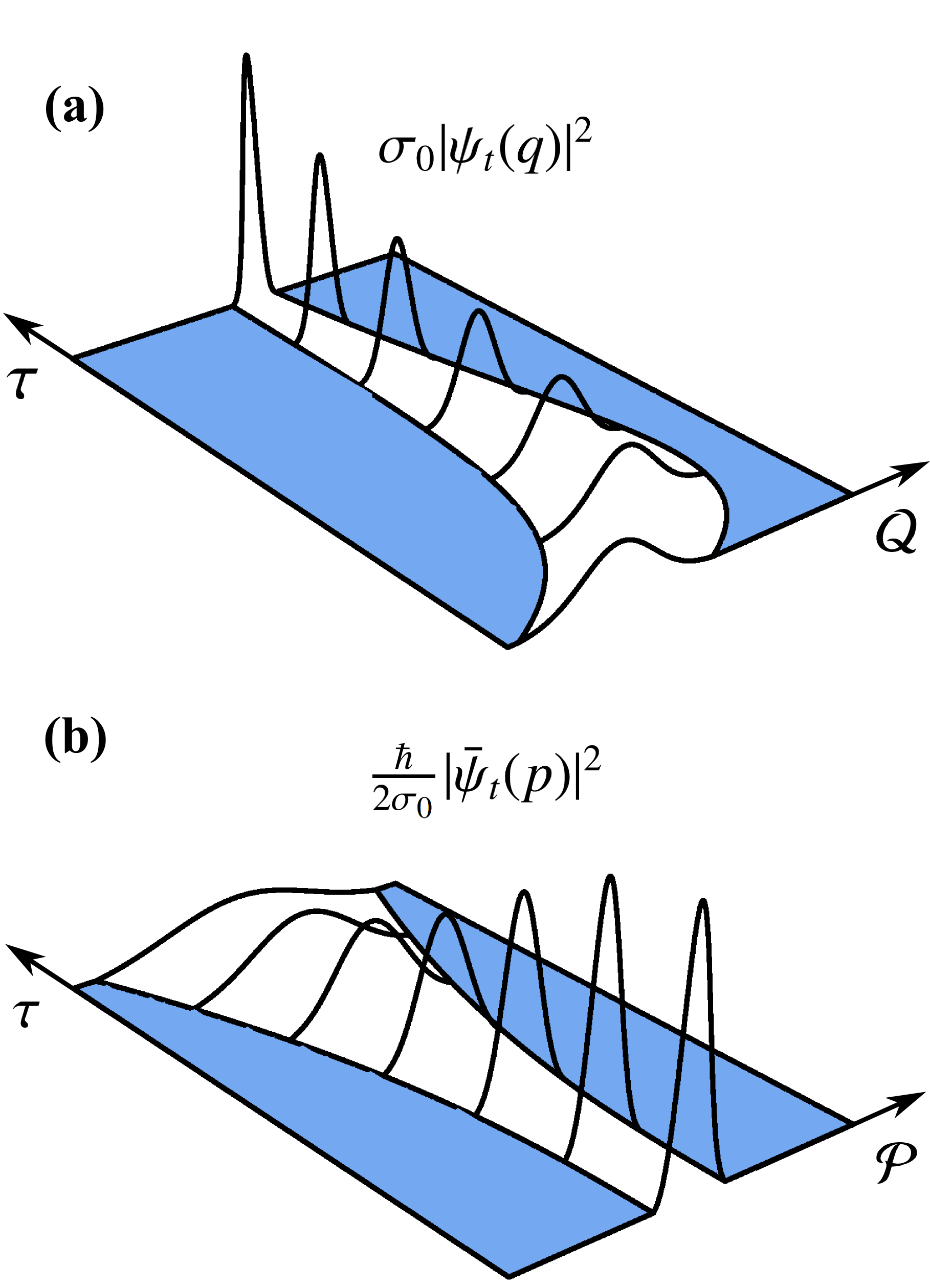}} 
\caption{Simulations for the probability densities in terms of the dimensionless quantities $\mathcal{Q}=q/\sigma_0$, $\mathcal{P}=2\sigma_0p/\hbar$, $\epsilon=k\sigma_0^2/(\hbar\lambda)=1.0$, $\tau_E=\lambda t_E=3.0$, $\zeta=(1-2\epsilon/\tau_E)^{1/2}=1/\sqrt{3}$, and $\tau=\lambda t$. The time interval between contiguous solid black lines is $\tau_{\max}/6$. The auxiliary blue basis defines a contour plot at the height $0.01$. (a) Scaled probability density $\sigma_0|\psi_t(q)|^2$ as a function of $\mathcal{Q}\in [-5.0,5.0]$ and $\tau \in [0,\tau_{\max}]$, with $\tau_{\max}=3\tau_E/2$, for $\mathcal{Q}_0=q_0/\sigma_0=-2.0$  and $\mathcal{P}_0=2\sigma_0 p_0/\hbar=20$. (b) Scaled probability density $\tfrac{\hbar}{2\sigma_0}|\bar{\psi}_t(p)|^2$ as a function of the canonical momentum $\mathcal{P}\in[-13,13]$ and $\tau\in [0,\tau_{\max}]$, with $\tau_{\max}=\tau_E/3$, for $\mathcal{Q}_0=\mathcal{P}_0=0$.}  
\label{fig2}
\end{figure}

The quantum CK model, \eqref{QCKH}, has a long history of conceptual discussions~\cite{Tartaglia1977,Greenberger1979,Greenberger1979a,Otero1984}, applications~\cite{Schuch1997,Schuch1999,Brown1991,Sun1995}, and derivations of analytical solutions~\cite{Dekker1981,Exner1983,Baskoutas1993,Pedrosa1997,Um2002,Razavy2006,Pepore2006}. Here we adopt the method developed in Ref.~\cite{Cheng1988}, which is particularly convenient for our purposes because it offers a formal solution for $i\hbar\,\partial_tU_t=H_tU_t$ in terms of the time-evolution operator $U_t$ in cases where the Hamiltonian can be written as
\be 
H_t=a_t^+\,J_++a_t^0\,J_0+a_t^-\,J_-, 
\label{HJ}
\ee 
where $J_{\pm,0}$ form the SU(2) Lie algebra characterized by $[J_+,J_-]=2J_0$ and $[J_0,J_{\pm}]=\pm J_{\pm}$, and $a_t^{\pm,0}$ are arbitrary functions of time. The identification of Hamiltonians \eqref{QCKH} and \eqref{HJ} is done via 
\be\begin{array}{lll}
\displaystyle a_t^+=\hbar\,\kappa_t, &\qquad& \displaystyle J_+=\frac{Q^2}{2\hbar}, \\ \\
\displaystyle a_t^0=0, && \displaystyle J_0=\frac{i\{Q,P\}}{4\hbar}, \\ \\
\displaystyle a_t^-=\frac{\hbar}{\mu_t}, && \displaystyle J_-=\frac{P^2}{2\hbar}, 
\end{array}\ee 
where $\mu_t=m\,e^{2\tau}$, $\kappa_t=k\,e^{2\tau}$, and $\{Q,P\}=QP+PQ$. The method then allows one to write
\be 
U_t=e^{i c_t^+J_+}\,e^{c_t^0J_0}\,e^{i c_t^-J_-},
\label{U_t}
\ee 
with time-dependent coefficients given by
\be 
c_t^+=\mu_t\frac{\dot{u}_t}{u_t},\quad c_t^0=-2\ln{\left(\frac{u_t}{u_0}\right)}, \quad c_t^-=-u_0^2\int_0^t\frac{dt'}{\mu_{t'}u_{t'}^2},
\ee 
with $c_0^{\pm,0}=0$ and $\ddot{u}_t+\frac{\dot{\mu}_t}{\mu_t}\dot{u}_t+\frac{\kappa_t}{\mu_t}u_t=0$, such that $u_0\neq 0$ and $\dot{u}_0=0$. With the explicit expressions for $\mu_t$ and $\kappa_t$ we find $\ddot{u}_t+2\lambda\dot{u}_t+\omega^2 u_t=0$, which shows that the quantum solution encapsulates the classical trajectory. Taking $u_0=1$ and $\dot{u}_0=0$ as initial conditions, we obtain the particular solution
\be 
u_t=e^{-\tau}\left[\cosh{(\zeta \tau)}+\frac{\sinh{(\zeta\tau)}}{\zeta} \right],
\ee 
which leads to
\begin{subequations}
\beq 
c_t^+&=&-\frac{k}{\lambda}\left[\frac{e^{2\tau}}{1+\zeta\coth{(\zeta\tau)}}\right], \\
c_t^0&=&2\ln{\left[\frac{e^{\tau}}{\cosh{(\zeta\tau)+\frac{\sinh{(\zeta\tau)}}{\zeta}}}\right]}, \\
c_t^-&=&-\frac{1}{m\lambda}\left[\frac{1}{1+\zeta\coth{(\zeta\tau)}}\right].
\eeq \label{Sol_c} 
\end{subequations}
Note that such solutions do not hold for $\lambda=0$. Having constructed the solution for the evolution operator, \eqref{U_t}, we can proceed with the calculation of the wave function. Since $J_0=\tfrac{1}{4}+\tfrac{iQP}{2\hbar}$ and $\bra{q}\exp{\left[(i c_t^0/2\hbar) QP\right]}\ket{\Theta}=e^{(c_t^0/2) q\partial_q}\Theta(q)=\Theta(e^{c_t^0/2}q)$, we find
\beq 
\psi_t(q)&=&\bra{q}U_t\ket{\psi_0}=e^{i c_t^+q^2/2\hbar}\,e^{c_t^0/4}\phi\left(e^{c_t^0/2}q,T_t\right),
\label{psi(q,t)}
\eeq 
where $\phi\left(q,T_t\right):=\bra{q}e^{ic_t^-J_-}\ket{\psi_0}$. From $ic_t^-J_-=-iP^2T_t/(2m\hbar)$ with
\be 
\lambda\, T_t:=\frac{1}{1+\zeta\coth{(\zeta\tau)}},
\label{T_t}
\ee 
we see that the solution for $\phi(q,T_t)$ can be obtained from the problem of a free particle evolving during a time interval $T_t$ (which is a monotonically increasing function of the dimensionless time $\tau$). Thus, taking the standard free-particle solution for a Gaussian probability density, we return to Eq. \eqref{psi(q,t)} to obtain, finally,
\be 
\big|\psi_t(q)\big|^2=\frac{\exp{\left[-\frac{\left(q-e^{-c_t^0/2}q_{T_t}\right)^2}{2(\Delta q_t)^2}\right]}}{\sqrt{2\pi(\Delta q_t)^2}}, \qquad\Delta q_t=\sigma_0\,\alpha_t\,e^{-c_t^0/2},
\label{psi2q}
\ee 
$\alpha_t=[1+(T_t/t_E)^2]^{1/2}$, $q_{T_t}=q_0+p_0T_t/m$, and $t_E=2m\sigma_0^2/\hbar$ (the Ehrenfest time). This solution presumes that at $t=0$ one has a Gaussian wave packet with mean values $(q_0,p_0)$ and uncertainties $\left(\sigma_0,\tfrac{\hbar}{2\sigma_0} \right)$. The centroid evolves in time according to $\langle Q\rangle_t=\exp{\big(-c_t^0/2\big)}\left(q_0+p_0T_t/m \right)$, with velocity $\langle V\rangle_t=d\langle Q\rangle_t/dt$. Since $c_t^0\simeq 2(1-\zeta)\tau$ and $\lambda T_t\simeq (1+\zeta)^{-1}$ for $t\to \infty$ and $\zeta\in(0,1)$, one has $\langle Q\rangle_t\simeq q_0e^{-(1-\zeta)\tau}$ and $\langle V\rangle_t\simeq -\lambda q_0 e^{-(1-\zeta)\tau}$, both vanishing for long times. Therefore, as time increases, the center of the wave packet starts to move as a free particle but inevitably goes to the origin of the coordinate system while the width rapidly diminishes. This dynamics is illustrated in Fig. \ref{fig2}(a).

As previously shown, the irreality of a Gaussian state depends only on its width, which, in the present case, is fully insensitive to the initial conditions $(q_0,p_0)$ of the wave packet. In addition, we can always find an inertial coordinate system for which the initial wave packet is at rest. Thus, for simplicity, henceforth we take $q_0=p_0=0$. With this simplification, the free-particle solution in the position representation reads $\phi(q,t)=(2\pi\sigma_t)^{1/4}\exp\left[-\left(1-i\,t/t_E\right)\,q^2/(4\sigma_t^2)\right]$, with $\sigma_t^2=\sigma_0^2[1+(t/t_E)^2]$. Plugging this result into \eqref{psi(q,t)} and taking the Fourier transform leads to $\bar{\psi}(p,t)$, whose squared modulus reads
\be 
\left|\bar{\psi}_t(p)\right|^2=\frac{\exp\left[-\frac{p^2}{2(\Delta p_t)^2}\right]}{\sqrt{2\pi(\Delta p_t)^2}}, \qquad \Delta p_t=\frac{\hbar}{2\sigma_0}\,\beta_t\,e^{c_t^0/2},
\label{psi2p}
\ee 
with $\beta_t=[1+4\,e^{-c_t^0}f_t]^{1/2}$, $f_t=\chi_t\left[(T_t/t_E)+\chi_t\left(\Delta q_{T_t}/\sigma_0\right)^2 \right]$, and $\chi_t=c_t^+\,\sigma_0^2/\hbar$. In Fig. \ref{fig2}(b), a simulation is presented for the time evolution of this probability distribution. It is immediately seen from solutions \eqref{psi2q} and $\eqref{psi2p}$ that Heisenberg's principle is always satisfied and that the asymptotic behaviors $\Delta q_t\simeq e^{-c_t^0/2}$ and $\Delta p_t\simeq e^{c_t^0/2}$ point to a full localization of the particle at $q=0$ and a complete delocalization of its momentum. All these results are in qualitative consonance with the classical results presented in Table \ref{tab1}.

We are now ready to assess the irrealities associated with position, momentum, and velocity. As previously discussed, the adequacy of formula \eqref{IrrGaussian} is conditioned to the relations $\Delta_{q,p}\geqslant 1$, which require $\delta q\leqslant \sigma_0\,\alpha_t\,e^{-c_t^0/2}$ and $\delta p\leqslant \tfrac{\hbar}{2\sigma_0}\,\beta_t\,e^{c_t^0/2}$ to hold simultaneously, and validate the uncertainty relation~\eqref{URQP}. These inequalities impose an upper bound for time as a function of $\sigma_0$, in such a way that the greater $\sigma_0$ the greater the time domain within which the irrealities are valid nonnegative quantities and the discretized model applies. Once such inequalities are respected, we then have, via Eq. \eqref{IrrGaussian}, the results
\begin{subequations}
\beq
\mathfrak{I}(Q|\rho_t)&=&\ln\left[\sqrt{2\pi e}\,\tfrac{\sigma_0}{\delta q}\,\alpha_t\,e^{-c_t^0/2} \right], \\
\mathfrak{I}(P|\rho_t)&=&\ln\left[\sqrt{2\pi e}\,\tfrac{\hbar}{2\sigma_0\delta p}\,\beta_t\,e^{c_t^0/2} \right],
\eeq
\end{subequations}
which show that the irreality of position (momentum) is a monotonically decreasing (increasing) function of time for $\rho_t=\ket{\psi_t}\bra{\psi_t}$, with $\ket{\psi_t}$ being the Gaussian state whose wave function is given by Eq. \eqref{psi(q,t)}. Even though the above results depend on the resolutions $\delta q$ and $\delta p$, the irreality variation $\Delta\mathfrak{I}_t^R:=\mathfrak{I}(R|\rho_t)-\mathfrak{I}(R|\rho_0)$ of the observable $R\in\{Q,P\}$ does not. One has $\Delta\mathfrak{I}_t^Q=\ln(\alpha_t\,e^{-c_t^0/2})$ and $\Delta\mathfrak{I}_t^P=\ln(\beta_t\,e^{c_t^0/2})$, which yield
\be 
\Delta\mathfrak{I}_t^Q+\Delta\mathfrak{I}_t^P=\ln{\big(\alpha_t\,\beta_t\big) }\stackrel{(t\to \infty)}{\longrightarrow} 2\,\zeta\, \tau.
\ee 
Therefore, while the position irreality is rapidly suppressed, the mean production rate $\Delta\mathfrak{I}_t^P/\Delta t$ of momentum irreality equals the constant $2\zeta\lambda$ already for times of the order of $t_E$. 

Although the behavior $\mathfrak{I}(Q|\rho_{\infty})\to 0$ is consistent with the notion of rest, the fact that $\mathfrak{I}(P|\rho_{\infty})\to\infty$ might, in principle, not be. However, as previously realized for the classical CK model, the correct observable to look at is the velocity. Using Heisenberg's equations, we find 
\be 
V_{H}:=\frac{dQ_H}{dt}=\frac{[Q_{H},H_H]}{i\hbar}=\frac{P_H}{m}\,e^{-2\tau}.
\ee 
It follows that $\langle V^n\rangle_t=\text{Tr}[\rho_0V_H^n]=\text{Tr}[\rho_0P_H^n]\tfrac{e^{-2\tau}}{m}=\langle P^n\rangle_t\tfrac{e^{-2\tau}}{m}$, for $n\geqslant 1\in\mathbb{N}$. We then find 
\be 
\Delta v_t:=\sqrt{\langle V^2\rangle_t-\langle V\rangle_t^2}=\frac{\Delta p_t}{m}\,e^{-2\tau}=\frac{\hbar}{2\sigma_0}\frac{\beta_t}{m}e^{-2\tau}e^{c_t^0/2}.
\ee 
Since $c_t^0\simeq 2\tau (1-\zeta)$ and $\beta_t\simeq e^{2\zeta\tau}$ for $t\to\infty$, in the overdamped regime ($0<\zeta< 1\in\mathbb{R}$), it is clear that $\Delta v_t\simeq e^{-(1-\zeta)\tau}$, which rapidly goes to 0 with time. Because the probability distribution for velocity is also Gaussian, we have $\mathfrak{I}(V|\rho_t)=\ln (\sqrt{2\pi e}\,\tfrac{\Delta v_t}{\delta v})$, with $\delta v= \delta p/m$. From the above estimates, it is easy to conclude that $\mathfrak{I}(V|\rho_t)$ will rapidly vanish as well. Therefore, we have found a framework where
\be 
\lim_{t\to \infty}\Big(\langle Q\rangle_t,\langle V\rangle_t,\mathfrak{I}(Q|\rho_t),\mathfrak{I}(V|\rho_t) \Big)=\boldsymbol{0},
\label{Qrest}
\ee 
meaning that not only are the average position and velocity consistent with the classical notion of rest, but also their respective elements of reality. It is noteworthy that the generality of the relation $\Delta v=\Delta p/m$ straightforwardly implies that $\Delta q\,\Delta p=m\Delta q\,\Delta v\geqslant \hbar/2$, which allows for the simultaneous emergence of reality for position and velocity ($\Delta q=\Delta v\to 0$) for heavy particles ($m\to \infty$), in harmonic coexistence with Heisenberg's uncertainty principle.

In trying to apply the present model to dynamically describe a position measurement, an important drawback is found: the asymptotic position of the particle, here interpreted as representative of the measurement outcome, is always 0. This does not reproduce the random aspect of measurement outcomes for a particle prepared in a Gaussian state. A possible way out of this difficulty is to consider that the above description actually refers to the position of the particle relative to a pointer, not to the laboratory reference frame. That is, let us hereafter assume that $q=x-x_{\wp}$ (a relative coordinate), where $x$ and $x_{\wp}$ denote the positions of the particle and the pointer $\wp$ relative to the laboratory. If we consider that the particle and the pointer form a perennial closed system, then we can assume that the center of mass $(\cm)$ remains uncorrelated with the relative coordinate for all times. Thus, the joint state of the system can be written as $\ket{\Psi_t}=\ket{\varphi_t}\otimes\ket{\psi_t}$, where $\ket{\psi_t}$ is the solution we just obtained above, adapted with the replacement $m\to \mu=m m_{\wp}/(m+m_{\wp})$, and $\ket{\varphi_t}$ is the state associated with the center of mass. For simplicity, in what follows we assume that $\varphi_t(x^{\cm})$ is the Gaussian solution of a free particle with $\langle X^{\cm}\rangle_t=\langle P^{\cm}\rangle_t=0$, $\Delta x^{\cm}_t=\sigma_{\cm}[1+(t/t_E^{\cm})^2]^{1/2}$, and $\Delta p^{\cm}_t=\hbar/(2\sigma_{\cm})$, where $t^{\cm}_E=2M\sigma_{\cm}^2/\hbar$ and $M=m+m_{\wp}$. We then have $\ket{\Psi_t}=\int\int dx^{\cm}dq\,\varphi_t(x^{\cm})\,\psi_t(q)\ket{x^{\cm}}\ket{q}$, with $\psi_t(q)$ given by Eq. \eqref{psi(q,t)}. Employing the traditional transformations from laboratory coordinates to center of mass plus relative coordinates, $x^{\cm}=\frac{m x+m_{\wp}x_{\wp}}{M}$ and $q=x-x_{\wp}$, we conceive the inverse map
\be 
\ket{x^{\cm}}\ket{q}\mapsto \ket{x^{\cm}+\tfrac{m_{\wp}}{M}q}\ket{x^{\cm}-\tfrac{m}{M}q}
\ee 
to link every state in $\mathcal{H}_{\cm}\otimes\mathcal{H}_{q}$ with one in $\mathcal{H}_{x}\otimes\mathcal{H}_{x_{\wp}}$. Using this map and performing a change of dummy variables, we rewrite the joint state as 
\be 
\ket{\Psi_t}=\int\int dx\,dx_{\wp}\,\varphi_t\left(\tfrac{m x+m_{\wp}x_{\wp}}{M}\right)\psi_t(x-x_{\wp})\,\ket{x}\ket{x_{\wp}}.
\ee 
As we learned from the discussion associated with Eq.~\eqref{psi2q}, as time passes $|\psi_t(x-x_{\wp})|$ approaches the Dirac delta function $\delta (x-x_{\wp})$. It follows that $\ket{\Psi_{\infty}}\simeq \int dx\,\varphi_{\infty}(x)\,\ket{x}\ket{x}$, which is a nonnormalizable estimate indicating that the asymptotic state is maximally entangled. A detailed computation of the degree of entanglement $E$ in the joint state can be obtained through the linear entropy of the reduced state, $E_t=1-{\Pi}[\rho_t^x]$, where $\Pi[\varrho]:=\text{Tr}(\varrho^2)$ is the purity of $\varrho$ and $\rho_t^x=\text{Tr}_{x_{\wp}}\ket{\Psi_t}\bra{\Psi_t}$ is the reduced state. Direct calculation yields
\be 
\Pi^2[\rho_t^x]=\frac{\gamma_t^2}{\left(1+\mathfrak{m}^2\,\gamma_t^2\right)\left(1+\mathfrak{m}_{\wp}^2\,\gamma_t^2\right)},  
\ee 
where $\gamma_t=\Delta q_t/\Delta x^{\cm}_t$, $\mathfrak{m}=m/M$, and $\mathfrak{m}_{\wp}=m_{\wp}/M$. Now, since $\Delta q_{\infty}\to 0$ while $\Delta x^{\cm}_{\infty}\to\infty$, one finds $\Pi[\rho_{\infty}^x]\to 0$ and $E_{\infty}\to 1$, which proves that entanglement eventually reaches its maximum for an overdamped dynamics. We see, therefore, for this two-body model, that the emergence of realism for the particle in the pointer reference frame is signalized, in the laboratory reference frame, as the creation of maximal quantum correlations. As thoroughly argued in Ref.~\cite{Dieguez2018a}, since the position of the particle is never directly accessed in any experiment---in fact, the observer only interacts with the pointer---we can trace out the corresponding subspace to obtain
\be 
\rho_{\infty}^{x_{\wp}}\simeq \int dx_{\wp}\left|\varphi_{\infty}(x_{\wp})\right|^2\ket{x_{\wp}}\bra{x_{\wp}}.
\ee 
Clearly, the asymptotic reduced state of the pointer is a fully incoherent statistical mixture, whose discretized version satisfies $\Phi_{X_{\wp}}(\rho_{\infty}^{x_{\wp}})=\rho_{\infty}^{x_{\wp}}$, which then implies an element of reality for the pointer position, that is, $\mathfrak{I}(X_{\wp}|\rho_{\infty}^{x_{\wp}})=0$. Therefore, given the inevitable discard of the particle position, the pointer position is certain to be in a real state. In this sense, the measurement problem dissipates.

\section{Summary}

By explicitly presenting a formalism through which one can quantify the degree of irreality associated with a continuous variable for a given quantum state, this work extends the approach recently put forward by Bilobran and Angelo~\cite{Bilobran2015}, which allows one to make inferences about realism in a quantitative fashion. As the first contribution, we derived the uncertainty, relation~\eqref{UR}, which indicates a lower bound for the total amount of irreality one can simultaneously set to arbitrary observables $A$ and $A'$ acting on $\mathcal{H_A}$ by preparing a quantum state. We then showed how to consistently discretize the position and momentum representations in terms of operational resolutions $\delta q$ and $\delta p$, which define the space dimension $L=(\xi-1)/2$ of the discretized model, with $\xi=2\pi\hbar/(\delta q \delta p)$. As expected, the continuous-variable formalism is fully retrieved as $\xi\to \infty$. With this strategy, we succeeded in explicitly computing the irreality of position and momentum for Gaussian states and, in agreement with inequality,~\eqref{UR}, deriving a position-momentum uncertainty relation [see inequality~\eqref{URQP}]. This result points out that the classical notion of a simultaneous position-momentum realism is forbidden by quantum mechanics in general.

As an application of the presently developed formalism, we demonstrated by example how the classical notion of rest can emerge from quantum mechanics. Using the CK model, we studied an effective dynamics whereby the mechanical energy of a harmonic oscillator is entirely dissipated into a reservoir while the system state remains pure. Even though the uncertainty and the irreality of the canonical momentum exponentially increase with time, it is shown that both the mean position and the mean velocity of the particle simultaneously go to 0 along with their respective irrealities [see Eq.~\eqref{Qrest}]. This is the expression of quantum rest, which occurs in full consistency with Heisenberg's uncertainty principle. Finally, applying the CK model to a two-body system, we showed that the dynamical emergence of reality for the particle from the perspective of the pointer manifests, in the laboratory reference frame, as the creation of maximum entanglement between the parts. Following Ref.~\cite{Dieguez2018a}, this result points to a solution for the measurement problem.

The techniques developed here, along with the quantifier introduced in Ref.~\cite{Bilobran2015}, constitute well- defined tools for the characterization of realism in a quantitative way. This may eventually be useful in several contexts involving spatial degrees of freedom, as, for instance, in foundational and applied studies related to arenas such as optomechanics, Stern-Gerlach experiments, Bell tests, quantum random walks, double-slit experiments, and quantum gravity, among others.

\section*{Acknowledgments}

The authors acknowledge support from CNPq and from the National Institute for Science and Technology of Quantum Information (INCT-IQ/CNPq, Brazil).



\begin{thebibliography}{80}

\expandafter\ifx\csname natexlab\endcsname\relax\def\natexlab#1{#1}\fi
\expandafter\ifx\csname bibnamefont\endcsname\relax
  \def\bibnamefont#1{#1}\fi
\expandafter\ifx\csname bibfnamefont\endcsname\relax
  \def\bibfnamefont#1{#1}\fi
\expandafter\ifx\csname citenamefont\endcsname\relax
  \def\citenamefont#1{#1}\fi
\expandafter\ifx\csname url\endcsname\relax
  \def\url#1{\texttt{#1}}\fi
\expandafter\ifx\csname urlprefix\endcsname\relax\def\urlprefix{URL }\fi
\providecommand{\bibinfo}[2]{#2}
\providecommand{\eprint}[2][]{\url{#2}}

\bibitem[{\citenamefont{Tonomura et al}(1989)}]{Tonomura1989}
\bibinfo{author}{A.~Tonomura}, \bibinfo{author}{J.~Endo}, \bibinfo{author}{T.~Matsuda}, \bibinfo{author}{T.~Kawasaki}, and \bibinfo{author}{H.~Ezawa}, 
  \bibinfo{title}{Demonstration of single-electron buildup of an interference pattern}, 
  \bibinfo{journal}{Am. J. Phys.} \bibinfo{volume}{{\bf 57}}, 
  \bibinfo{pages}{117} (\bibinfo{year}{1989}).

\bibitem{Arndt1999}
\bibinfo{author}{M.~Arndt}, \bibinfo{author}{O.~Nairz}, \bibinfo{author}{J.~Vos-Andreae}, \bibinfo{author}{C.~Keller}, \bibinfo{author}{G.~van~der Zouw}, and \bibinfo{author}{A.~Zeilinger},
  \bibinfo{title}{Wave-particle duality of C60 molecules},
  \bibinfo{journal}{Nature} \bibinfo{volume}{{\bf 401}}, 
  \bibinfo{pages}{680} (\bibinfo{year}{1999}).
  
\bibitem{Schrodinger1935}
\bibinfo{author}{E.~Schr{\"o}dinger}, 
  \bibinfo{title}{Die gegenw{\"a}rtige situation in der quantenmechanik}, 
  \bibinfo{journal}{Naturwissenschaften} \bibinfo{volume}{{\bf 23}},
  \bibinfo{pages}{807} (\bibinfo{year}{1935}).

\bibitem{Einstein1935}
\bibinfo{author}{A.~Einstein}, \bibinfo{author}{B.~Podolsky}, and \bibinfo{author}{N.~Rosen}, 
  \bibinfo{title}{Can quantum-mechanical description of physical reality be considered complete?}
  \bibinfo{journal}{Phys. Rev.} \bibinfo{volume}{{\bf 47}},
   \bibinfo{pages}{777} (\bibinfo{year}{1935}).

\bibitem{Bell1964}
\bibinfo{author}{J.~S. Bell}, 
  \bibinfo{title}{On the Einstein-Podolsky-Rosen paradox}, 
  \bibinfo{journal}{Physics} \bibinfo{volume}{{\bf 1}},
  \bibinfo{pages}{195} (\bibinfo{year}{1964}).

\bibitem{Bohm1952a}
\bibinfo{author}{D.~Bohm}, 
  \bibinfo{title}{A suggested interpretation of the quantum theory in terms of ``hidden'' variables. I}, 
  \bibinfo{journal}{Phys. Rev.} \bibinfo{volume}{{\bf 85}}, 
  \bibinfo{pages}{166} (\bibinfo{year}{1952}{\natexlab{a}}).

\bibitem{Bohm1952b}
\bibinfo{author}{D.~Bohm}, 
  \bibinfo{title}{A suggested interpretation of the quantum theory in terms of ``hidden'' variables. II}, 
  \bibinfo{journal}{Phys. Rev.} \bibinfo{volume}{{\bf 85}}, 
  \bibinfo{pages}{180} (\bibinfo{year}{1952}{\natexlab{b}}).
  
\bibitem{Hensen2015}
\bibinfo{author}{B.~Hensen}, \bibinfo{author}{H.~Bernien}, \bibinfo{author}{A.~E. Dr\'{e}au}, \bibinfo{author}{A.~Reiserer}, \bibinfo{author}{N.~Kalb}, \bibinfo{author}{M.~S. Blok}, \bibinfo{author}{J.~Ruitenberg}, \bibinfo{author}{R.~F.~L. Vermeulen}, \bibinfo{author}{R.~N. Schouten}, \bibinfo{author}{C.~Abellán}, \bibinfo{author}{W.~Amaya}, \bibinfo{author}{V.~Pruneri}, \bibinfo{author}{M.~W. Mitchell}, \bibinfo{author}{M.~Markham}, \bibinfo{author}{D.~J. Twitchen}, \bibinfo{author}{D.~Elkouss}, \bibinfo{author}{S.~Wehner}, \bibinfo{author}{T.~H. Taminiau}, and \bibinfo{author}{R.~Hanson}, 
  \bibinfo{title}{Loophole-free Bell inequality violation using electron spins separated by 1.3 kilometres},
  \bibinfo{journal}{Nature} \bibinfo{volume}{{\bf 526}}, 
  \bibinfo{pages}{682} (\bibinfo{year}{2015}).

\bibitem{Giustina2015}
\bibinfo{author}{M.~Giustina}, \bibinfo{author}{M.~A.~M. Versteegh}, \bibinfo{author}{S.~Wengerowsky}, \bibinfo{author}{J.~Handsteiner}, \bibinfo{author}{A.~Hochrainer}, \bibinfo{author}{K.~Phelan}, \bibinfo{author}{F.~Steinlechner}, \bibinfo{author}{J.~Kofler}, \bibinfo{author}{J.-A. Larsson}, \bibinfo{author}{C.~Abell\'an}, \bibinfo{author}{W.~Amaya}, \bibinfo{author}{V.~Pruneri}, \bibinfo{author}{M.~W. Mitchell}, \bibinfo{author}{J.~Beyer}, \bibinfo{author}{T.~Gerrits}, \bibinfo{author}{A.~E. Lita}, \bibinfo{author}{L.~K. Shalm}, \bibinfo{author}{S.~W. Nam}, \bibinfo{author}{T.~Scheidl}, \bibinfo{author}{R.~Ursin}, \bibinfo{author}{B.~Wittmann}, and \bibinfo{author}{A.~Zeilinger}, 
  \bibinfo{title}{Significant-loophole-free test of Bell's theorem with entangled photons}, 
  \bibinfo{journal}{Phys. Rev. Lett.} \bibinfo{volume}{{\bf 115}}
  \bibinfo{pages}{250401} (\bibinfo{year}{2015}).
  
\bibitem{Shalm2015}
\bibinfo{author}{L.~K. Shalm}, \bibinfo{author}{E.~Meyer-Scott}, \bibinfo{author}{B.~G. Christensen}, \bibinfo{author}{P.~Bierhorst}, \bibinfo{author}{M.~A. Wayne}, \bibinfo{author}{M.~J. Stevens}, \bibinfo{author}{T.~Gerrits}, \bibinfo{author}{S.~Glancy}, \bibinfo{author}{D.~R. Hamel}, \bibinfo{author}{M.~S. Allman}, \bibinfo{author}{K.~J. Coakley}, \bibinfo{author}{S.~D. Dyer}, \bibinfo{author}{C.~Hodge}, \bibinfo{author}{A.~E. Lita}, \bibinfo{author}{V.~B. Verma}, \bibinfo{author}{C.~Lambrocco}, \bibinfo{author}{E.~Tortorici}, \bibinfo{author}{A.~L. Migdall}, \bibinfo{author}{Y.~Zhang}, \bibinfo{author}{D.~R. Kumor}, \bibinfo{author}{W.~H. Farr}, \bibinfo{author}{F.~Marsili}, \bibinfo{author}{M.~D. Shaw}, \bibinfo{author}{J.~A. Stern}, \bibinfo{author}{C.~Abell\'an}, \bibinfo{author}{W.~Amaya}, \bibinfo{author}{V.~Pruneri}, \bibinfo{author}{T.~Jennewein}, \bibinfo{author}{M.~W. Mitchell}, \bibinfo{author}{P.~G. Kwiat}, \bibinfo{author}{J.~C. Bienfang}, \bibinfo{author}{R.~P. Mirin}, \bibinfo{author}{E.~Knill}, and \bibinfo{author}{S.~W. Nam},
  \bibinfo{title}{Strong loophole-free test of local realism},
  \bibinfo{journal}{Phys. Rev. Lett.} \bibinfo{volume}{{\bf 115}},
  \bibinfo{pages}{250402--250411} (\bibinfo{year}{2015}). 

\bibitem{Hensen2016}
\bibinfo{author}{B.~Hensen}, \bibinfo{author}{N.~Kalb}, \bibinfo{author}{M.~S.~Blok}, \bibinfo{author}{A.~E. Dr\'{e}au}, \bibinfo{author}{A.~Reiserer}, \bibinfo{author}{R.~F.~L. Vermeulen}, \bibinfo{author}{R.~N. Schouten}, \bibinfo{author}{M.~Markham}, \bibinfo{author}{D.~J. Twitchen}, \bibinfo{author}{K.~Goodenough}, \bibinfo{author}{D.~Elkouss}, \bibinfo{author}{S.~Wehner}, \bibinfo{author}{T.~H. Taminiau}, and \bibinfo{author}{R.~Hanson}, 
  \bibinfo{title}{Loophole-free Bell test using electron spins in diamond: second experiment and additional analysis},
  \bibinfo{journal}{Sci. Rep.} \bibinfo{volume}{{\bf 6}},
   \bibinfo{pages}{30289} (\bibinfo{year}{2016}).

\bibitem{Rauch2018}
\bibinfo{author}{D.~Rauch}, \bibinfo{author}{J.~Handsteiner}, \bibinfo{author}{A.~Hochrainer}, \bibinfo{author}{J.~Gallicchio}, \bibinfo{author}{A.~S. Friedman}, \bibinfo{author}{C.~Leung}, \bibinfo{author}{B.~Liu}, \bibinfo{author}{L.~Bulla}, \bibinfo{author}{S.~Ecker}, \bibinfo{author}{F.~Steinlechner}, \bibinfo{author}{R.~Ursin}, \bibinfo{author}{B.~Hu}, \bibinfo{author}{D.~Leon}, \bibinfo{author}{C.~Benn}, \bibinfo{author}{A.~Ghedina}, \bibinfo{author}{M.~Cecconi}, \bibinfo{author}{A.~H. Guth}, \bibinfo{author}{D.~I. Kaiser}, \bibinfo{author}{T.~Scheidl}, and \bibinfo{author}{A.~Zeilinger},
  \bibinfo{title}{Cosmic Bell test using random measurement settings from high-redshift quasars}, 
  \bibinfo{journal}{Phys. Rev. Lett.} \bibinfo{volume}{{\bf 121}}, 
  \bibinfo{pages}{080403} (\bibinfo{year}{2018}).

\bibitem{Li2018}
\bibinfo{author}{M.-H. Li}, \bibinfo{author}{C.~Wu}, \bibinfo{author}{Y.~Zhang}, \bibinfo{author}{W.-Z. Liu}, \bibinfo{author}{B.~Bai}, \bibinfo{author}{Y.~Liu}, \bibinfo{author}{W.~Zhang}, \bibinfo{author}{Q.~Zhao}, \bibinfo{author}{H.~Li}, \bibinfo{author}{Z.~Wang}, \bibinfo{author}{L.~You}, \bibinfo{author}{W.~J. Munro}, \bibinfo{author}{J.~Yin}, \bibinfo{author}{J.~Zhang}, \bibinfo{author}{C.-Z. Peng}, \bibinfo{author}{X.~Ma}, \bibinfo{author}{Q.~Zhang}, \bibinfo{author}{J.~Fan}, and \bibinfo{author}{J.-W. Pan}, 
  \bibinfo{title}{Test of local realism into the past without detection and locality loopholes},
  \bibinfo{journal}{Phys. Rev. Lett.} \bibinfo{volume}{{\bf 121}},
  \bibinfo{pages}{080404} (\bibinfo{year}{2018}).

\bibitem{Pusey2012}
\bibinfo{author}{M.~F. Pusey}, \bibinfo{author}{J.~Barrett}, and \bibinfo{author}{T.~Rudolph}, 
  \bibinfo{title}{On the reality of the quantum state}, 
  \bibinfo{journal}{Nat. Phys.} \bibinfo{volume}{{\bf 8}},
  \bibinfo{pages}{475} (\bibinfo{year}{2012}).

\bibitem{Lewis2012}
\bibinfo{author}{P.~G. Lewis}, \bibinfo{author}{D.~Jennings}, \bibinfo{author}{J.~Barrett}, and \bibinfo{author}{T.~Rudolph},
  \bibinfo{title}{Distinct quantum states can be compatible with a single state of reality}, 
  \bibinfo{journal}{Phys. Rev. Lett.} \bibinfo{volume}{{\bf 109}},
  \bibinfo{pages}{1} (\bibinfo{year}{2012}).

\bibitem{Colbeck2012}
\bibinfo{author}{R.~Colbeck} and \bibinfo{author}{R.~Renner}, 
  \bibinfo{title}{Is a system's wave function in one-to-one correspondence with its elements of reality?} 
  \bibinfo{journal}{Phys. Rev. Lett.} \bibinfo{volume}{{\bf 108}},
  \bibinfo{pages}{1} (\bibinfo{year}{2012}).

\bibitem{Hardy2013}
\bibinfo{author}{L.~Hardy}, 
  \bibinfo{title}{Are quantum states real?}
  \bibinfo{journal}{Int. J. Mod. Phys. B} \bibinfo{volume}{{\bf 27}}, 
  \bibinfo{pages}{1345012} (\bibinfo{year}{2013}).

\bibitem{Patra2013}
\bibinfo{author}{M.~K. Patra}, \bibinfo{author}{S.~Pironio}, and \bibinfo{author}{S.~Massar}, 
  \bibinfo{title}{No-go theorems for $\psi$-epistemic models based on a continuity assumption},
  \bibinfo{journal}{Phys. Rev. Lett.} \bibinfo{volume}{{\bf 111}}, 
  \bibinfo{pages}{1} (\bibinfo{year}{2013}).

\bibitem{Aaronson2013}
\bibinfo{author}{S.~Aaronson}, \bibinfo{author}{A.~Bouland}, \bibinfo{author}{L.~Chua}, and \bibinfo{author}{G.~Lowther},
  \bibinfo{title}{$\ensuremath{\psi}$-epistemic theories: The role of symmetry}, 
  \bibinfo{journal}{Phys. Rev. A} \bibinfo{volume}{{\bf 88}},
  \bibinfo{pages}{032111} (\bibinfo{year}{2013}).

\bibitem{Leifer2014}
\bibinfo{author}{M.~S. Leifer}, 
  \bibinfo{title}{$\ensuremath{\psi}$-epistemic models are exponentially bad at explaining the distinguishability of quantum states}, 
  \bibinfo{journal}{Phys. Rev. Lett.} \bibinfo{volume}{{\bf 112}},
   \bibinfo{pages}{160404} (\bibinfo{year}{2014}).

\bibitem{Barrett2014}
\bibinfo{author}{J.~Barrett}, \bibinfo{author}{E.~G. Cavalcanti}, \bibinfo{author}{R.~Lal}, and \bibinfo{author}{O.~J.~E. Maroney},
  \bibinfo{title}{No $\ensuremath{\psi}$-epistemic model can fully explain the indistinguishability of quantum states}, 
  \bibinfo{journal}{Phys. Rev. Lett.} \bibinfo{volume}{{\bf 112}}, 
  \bibinfo{pages}{250403} (\bibinfo{year}{2014}).

\bibitem{Branciard2014}
\bibinfo{author}{C.~Branciard}, 
  \bibinfo{title}{How $\ensuremath{\psi}$-epistemic models fail at explaining the indistinguishability of quantum states}, 
  \bibinfo{journal}{Phys. Rev. Lett.} \bibinfo{volume}{{\bf 113}}, 
  \bibinfo{pages}{020409} (\bibinfo{year}{2014}). 
 
\bibitem{Ringbauer2015}
\bibinfo{author}{M.~Ringbauer}, \bibinfo{author}{B.~Duffus}, \bibinfo{author}{C.~Branciard}, \bibinfo{author}{E.~G. Cavalcanti}, \bibinfo{author}{A.~G. White}, and \bibinfo{author}{A.~Fedrizzi},
  \bibinfo{title}{Measurements on the reality of the wavefunction},
  \bibinfo{journal}{Nat. Phys.} \bibinfo{volume}{{\bf 11}},
  \bibinfo{pages}{249} (\bibinfo{year}{2015}).
  
\bibitem{Ballentine1970}
\bibinfo{author}{L.~E. Ballentine}, 
  \bibinfo{title}{The statistical interpretation of quantum mechanics}, 
  \bibinfo{journal}{Rev. Mod. Phys.} \bibinfo{volume}{{\bf 42}},
  \bibinfo{pages}{358} (\bibinfo{year}{1970}).
  
\bibitem{Emerson2001}
\bibinfo{author}{J.~V. Emerson}, 
  \bibinfo{title}{{\it Quantum Chaos and Quantum-Classical Correspondence}} 
  (\bibinfo{publisher}{PhD dissertation, Simon Fraser Universit}, \bibinfo{year}{2001}).

\bibitem{Spekkens2007}
\bibinfo{author}{R.~W. Spekkens}, 
  \bibinfo{title}{Evidence for the epistemic view of quantum states: A toy theory}, 
  \bibinfo{journal}{Phys. Rev. A} \bibinfo{volume}{{\bf 75}}, 
  \bibinfo{pages}{032110} (\bibinfo{year}{2007}).

\bibitem{vanEnk2007}
\bibinfo{author}{S.~J. van Enk}, 
  \bibinfo{title}{A toy model for quantum mechanics}, 
  \bibinfo{journal}{Found. Phys.} \bibinfo{volume}{{\bf 37}},
  \bibinfo{pages}{1447} (\bibinfo{year}{2007}).
  
\bibitem{Harrigan2010}
\bibinfo{author}{N.~Harrigan} and \bibinfo{author}{R.~W. Spekkens},
  \bibinfo{title}{Einstein, incompleteness, and the epistemic view of quantum states}, 
  \bibinfo{journal}{Found. Phys.} \bibinfo{volume}{{\bf 40}}, 
  \bibinfo{pages}{125} (\bibinfo{year}{2010}).

\bibitem{Bartlett2012}
\bibinfo{author}{S.~D. Bartlett}, \bibinfo{author}{T.~Rudolph}, and \bibinfo{author}{R.~W. Spekkens}, 
  \bibinfo{title}{Reconstruction of Gaussian quantum mechanics from Liouville mechanics with an epistemic restriction},
  \bibinfo{journal}{Phys. Rev. A} \bibinfo{volume}{{\bf 86}}, 
  \bibinfo{pages}{012103} (\bibinfo{year}{2012}).

\bibitem{Spekkens2015}
\bibinfo{author}{R.~W. Spekkens}, in
  \bibinfo{title}{{\it Quantum Theory: Informational Foundations and Foils}, edited by G. Chiribella and R. W. Spekkens},
  (\bibinfo{publisher}{Springer, Dordrecht, Netherlands}, \bibinfo{year}{2015}).

\bibitem{Budiyono2017}
\bibinfo{author}{A.~Budiyono} and \bibinfo{author}{D.~Rohrlich},
  \bibinfo{title}{Quantum mechanics as classical statistical mechanics with an ontic extension and an epistemic restriction}, 
  \bibinfo{journal}{Nat. Commun.} \bibinfo{volume}{{\bf 8}}, 
   \bibinfo{pages}{1306} (\bibinfo{year}{2017}).

\bibitem{Zurek2009}
\bibinfo{author}{W.~H. Zurek}, 
  \bibinfo{title}{Quantum Darwinism},
  \bibinfo{journal}{Nat. Phys.} \bibinfo{volume}{{\bf 11}}, 
  \bibinfo{pages}{181} (\bibinfo{year}{2009}).

\bibitem{Burke2010}
\bibinfo{author}{A.~M. Burke}, \bibinfo{author}{R.~Akis}, \bibinfo{author}{T.~E. Day}, \bibinfo{author}{G.~Speyer}, \bibinfo{author}{D.~K. Ferry}, and \bibinfo{author}{B.~R. Bennett}, 
  \bibinfo{title}{Periodic scarred states in open quantum dots as evidence of quantum Darwinism}, 
  \bibinfo{journal}{Phys. Rev. Lett.} \bibinfo{volume}{{\bf 104}},
  \bibinfo{pages}{176801} (\bibinfo{year}{2010}).

\bibitem{Riedel2010}
\bibinfo{author}{C.~J. Riedel} and \bibinfo{author}{W.~H. Zurek},
  \bibinfo{title}{Quantum Darwinism in an everyday environment: Huge redundancy in scattered photons}, 
  \bibinfo{journal}{Phys. Rev. Lett.} \bibinfo{volume}{{\bf 105}},
  \bibinfo{pages}{020404} (\bibinfo{year}{2010}).
 
\bibitem{Horodecki2015}
\bibinfo{author}{R.~Horodecki}, \bibinfo{author}{J.~K. Korbicz}, and \bibinfo{author}{P.~Horodecki}, 
  \bibinfo{title}{Quantum origins of objectivity}, 
  \bibinfo{journal}{Phys. Rev. A} \bibinfo{volume}{{\bf 91}},
  \bibinfo{pages}{032122} (\bibinfo{year}{2015}).

\bibitem{Brandao2015}
\bibinfo{author}{F.~G. S.~L. Brand\~ao}, \bibinfo{author}{M.~Piani}, and \bibinfo{author}{P.~Horodecki}, 
  \bibinfo{title}{Generic emergence of classical features in quantum Darwinism}, 
  \bibinfo{journal}{Nat. Commun.} \bibinfo{volume}{{\bf 6}},
  \bibinfo{pages}{7908} (\bibinfo{year}{2015}).

\bibitem{Zeilinger1999}
\bibinfo{author}{A.~Zeilinger}, 
  \bibinfo{title}{A foundational principle for quantum mechanics}, 
  \bibinfo{journal}{Found. Phys.} \bibinfo{volume}{{\bf 29}},
  \bibinfo{pages}{631} (\bibinfo{year}{1999}).

\bibitem{Brukner1999}
\bibinfo{author}{\v{C}.~Brukner} and \bibinfo{author}{A.~Zeilinger}, 
  \bibinfo{title}{Operationally invariant information in quantum measurements},
  \bibinfo{journal}{Phys. Rev. Lett.} \bibinfo{volume}{{\bf 83}},
  \bibinfo{pages}{3354} (\bibinfo{year}{1999}).

\bibitem{Brukner2003}
\bibinfo{author}{\v{C}.~Brukner} and \bibinfo{author}{A.~Zeilinger}, in
  \bibinfo{title}{{\it Time, Quantum and Information}, edited by L. Castell and O. Ischebeck}, 
  (\bibinfo{publisher}{Springer, Dordrecht, Netherlands}, \bibinfo{year}{2015}).
  
\bibitem{Bohr1935}
\bibinfo{author}{N.~Bohr}, 
  \bibinfo{title}{Can quantum-mechanical description of physical reality be considered complete?} 
  \bibinfo{journal}{Phys. Rev.} \bibinfo{volume}{{\bf 48}},
  \bibinfo{pages}{696} (\bibinfo{year}{1935}).

\bibitem{Angelo2015}
\bibinfo{author}{R.~M. Angelo} and \bibinfo{author}{A.~D. Ribeiro},
  \bibinfo{title}{Wave{\textendash}particle duality: An information-based approach}, 
  \bibinfo{journal}{Found. Phys.} \bibinfo{volume}{{\bf 45}},
  \bibinfo{pages}{1407} (\bibinfo{year}{2015}).

\bibitem{Ruark1935}
\bibinfo{author}{A.~E. Ruark}, 
  \bibinfo{title}{Is the quantum-mechanical description of physical reality complete?} 
  \bibinfo{journal}{Phys. Rev.} \bibinfo{volume}{{\bf 48}}, 
  \bibinfo{pages}{466} (\bibinfo{year}{1935}).
 
\bibitem{Redhead1989}
\bibinfo{author}{M.~Redhead}, 
  \bibinfo{title}{{\it Incompleteness, Nolocality, and Realism}} 
  (\bibinfo{publisher}{Oxford University Press, Oxford, UK}, \bibinfo{year}{1989}).

\bibitem{Vaidman1996}
\bibinfo{author}{L.~Vaidman}, 
  \bibinfo{title}{Weak-measurement elements of reality}, 
  \bibinfo{journal}{Found. Phys.} \bibinfo{volume}{{\bf 26}},
  \bibinfo{pages}{895} (\bibinfo{year}{1996}).

\bibitem{Bilobran2015}
\bibinfo{author}{A.~L.~O. Bilobran} and \bibinfo{author}{R.~M. Angelo},
  \bibinfo{title}{A measure of physical reality}, 
  \bibinfo{journal}{Europhys. Lett.} \bibinfo{volume}{{\bf 112}},
  \bibinfo{pages}{40005} (\bibinfo{year}{2015}).

\bibitem{Gomes2018}
\bibinfo{author}{V.~S. Gomes} and \bibinfo{author}{R.~M. Angelo},
  \bibinfo{title}{Nonanomalous measure of realism-based nonlocality},
  \bibinfo{journal}{Phys. Rev. A} \bibinfo{volume}{{\bf 97}}, 
  \bibinfo{pages}{012123} (\bibinfo{year}{2018}).

\bibitem{Gomes2019}
\bibinfo{author}{V.~S. Gomes} and \bibinfo{author}{R.~M. Angelo},
  \bibinfo{title}{Resilience of realism-based nonlocality to local disturbance}, 
  \bibinfo{journal}{Phys. Rev. A} \bibinfo{volume}{{\bf 99}}
  \bibinfo{pages}{012109} (\bibinfo{year}{2019}).

\bibitem{Dieguez2018a}
\bibinfo{author}{P.~R. Dieguez} and \bibinfo{author}{R.~M. Angelo},
  \bibinfo{title}{Information-reality complementarity: The role of measurements and quantum reference frames}, 
  \bibinfo{journal}{Phys. Rev. A} \bibinfo{volume}{{\bf 97}},
  \bibinfo{pages}{022107} (\bibinfo{year}{2018}).
  
\bibitem{Mancino2018}
\bibinfo{author}{L.~Mancino}, \bibinfo{author}{M.~Sbroscia}, \bibinfo{author}{E.~Roccia}, \bibinfo{author}{I.~Gianani}, \bibinfo{author}{V.~Cimini}, \bibinfo{author}{M.~Paternostro}, and \bibinfo{author}{M.~Barbieri}, 
  \bibinfo{title}{Information-reality complementarity in photonic weak measurements},
  \bibinfo{journal}{Phys. Rev. A} \bibinfo{volume}{{\bf 97}},
  \bibinfo{pages}{062108} (\bibinfo{year}{2018}).
  
\bibitem{Nielsen2000}
\bibinfo{author}{M.~A. Nielsen} and \bibinfo{author}{I.~L. Chuang},
  \bibinfo{title}{{\it Quantum Computation and Quantum Information}}
  (\bibinfo{publisher}{Cambridge University Press, Cambridge, UK}, \bibinfo{year}{2000}).
  
\bibitem{Baumgratz2014}
\bibinfo{author}{T.~Baumgratz}, \bibinfo{author}{M.~Cramer}, and \bibinfo{author}{M.~B. Plenio}, 
  \bibinfo{title}{Quantifying coherence},
  \bibinfo{journal}{Phys. Rev. Lett.} \bibinfo{volume}{{\bf 113}},
  \bibinfo{pages}{140401} (\bibinfo{year}{2014}).
  
\bibitem{Rulli2011}
\bibinfo{author}{C.~C. Rulli} and \bibinfo{author}{M.~S. Sarandy}, 
  \bibinfo{title}{Global quantum discord in multipartite systems},
  \bibinfo{journal}{Phys. Rev. A} \bibinfo{volume}{{\bf 84}},
  \bibinfo{pages}{042109} (\bibinfo{year}{2011}).
  
\bibitem{Rudnicki2018}
\bibinfo{author}{L.~Rudnicki}, 
  \bibinfo{title}{Uncertainty-reality complementarity and entropic uncertainty relations},
  \bibinfo{journal}{J. Phys. A} \bibinfo{volume}{{\bf 51}},
  \bibinfo{pages}{504001} (\bibinfo{year}{2018}).

\bibitem{Sakurai2010}
\bibinfo{author}{J.~J. Sakurai} and \bibinfo{author}{J.~J. Napolitano},
  \bibinfo{title}{{\it Modern Quantum Mechanics}} 
  (\bibinfo{publisher}{Pearson}, San Francisco, \bibinfo{year}{2010}).

\bibitem{Birula2006}
\bibinfo{author}{I.~Bialynicki-Birula}, 
  \bibinfo{title}{Formulation of the uncertainty relations in terms of the R\'enyi entropies},
  \bibinfo{journal}{Phys. Rev. A} \bibinfo{volume}{{\bf 74}}, 
  \bibinfo{pages}{052101} (\bibinfo{year}{2006}).

\bibitem{Caldirola1941}
\bibinfo{author}{P.~Caldirola}, 
  \bibinfo{title}{Forze non conservative nella meccanica quantistica}, 
  \bibinfo{journal}{Il Nuovo Cimento} \bibinfo{volume}{{\bf 18}}, 
  \bibinfo{pages}{393} (\bibinfo{year}{1941}).
  
\bibitem{Kanai1948}
\bibinfo{author}{E.~Kanai}, 
  \bibinfo{title}{On the quantization of the dissipative systems}, 
  \bibinfo{journal}{Progr. Theor. Exp. Phys.} \bibinfo{volume}{{\bf 3}}, 
  \bibinfo{pages}{440} (\bibinfo{year}{1948}).

\bibitem{Tartaglia1977}
\bibinfo{author}{A.~Tartaglia}, 
  \bibinfo{title}{A canonical approach to the quantum problem on the motion of a particle in a viscous medium},
  \bibinfo{journal}{Lett. Nuovo Cimento} \bibinfo{volume}{{\bf 19}}, 
  \bibinfo{pages}{205} (\bibinfo{year}{1977}).

\bibitem{Greenberger1979}
\bibinfo{author}{D.~M. Greenberger}, 
  \bibinfo{title}{A new approach to the problem of dissipation in quantum mechanics}, 
  \bibinfo{journal}{J. Math. Phys.} \bibinfo{volume}{{\bf 20}},
  \bibinfo{pages}{771} (\bibinfo{year}{1979}).

\bibitem{Greenberger1979a}
\bibinfo{author}{D.~M. Greenberger}, 
  \bibinfo{title}{A critique of the major approaches to damping in quantum theory}, 
  \bibinfo{journal}{J. Math. Phys.} \bibinfo{volume}{{\bf 20}},
  \bibinfo{pages}{762} (\bibinfo{year}{1979}).

\bibitem{Otero1984}
\bibinfo{author}{D.~Otero}, \bibinfo{author}{A.~Plastino}, \bibinfo{author}{A.~Proto}, and \bibinfo{author}{G.~Zannoli},
  \bibinfo{title}{Quantal friction, nonlinear Hamiltonians, and information theory}, 
  \bibinfo{journal}{Z. Phys. A} \bibinfo{volume}{{\bf 316}}, 
  \bibinfo{pages}{323} (\bibinfo{year}{1984}).

\bibitem{Schuch1997}
\bibinfo{author}{D.~Schuch}, 
  \bibinfo{title}{Nonunitary connection between explicitly time-dependent and nonlinear approachesfor the description of dissipative quantum systems},    
  \bibinfo{journal}{Phys. Rev. A} \bibinfo{volume}{{\bf 55}},
  \bibinfo{pages}{935} (\bibinfo{year}{1997}).

\bibitem{Schuch1999}
\bibinfo{author}{D.~Schuch}, 
  \bibinfo{title}{Effective description of the dissipative interaction between simple model-systems and their environment},
  \bibinfo{journal}{Int. J. Quantum Chem.} \bibinfo{volume}{{\bf 72}}, 
  \bibinfo{pages}{537} (\bibinfo{year}{1999}).

\bibitem{Brown1991}
\bibinfo{author}{L.~S. Brown}, 
  \bibinfo{title}{Quantum motion in a Paul trap},
  \bibinfo{journal}{Phys. Rev. Lett.} \bibinfo{volume}{{\bf 66}}, 
  \bibinfo{pages}{527} (\bibinfo{year}{1991}).
  
\bibitem{Sun1995}
\bibinfo{author}{C.-P. Sun} and \bibinfo{author}{L.-H. Yu}, 
  \bibinfo{title}{Exact dynamics of a quantum dissipative system in a constant external field},
  \bibinfo{journal}{Phys. Rev. A} \bibinfo{volume}{{\bf 51}},
   \bibinfo{pages}{1845} (\bibinfo{year}{1995}).
 
\bibitem{Dekker1981}
\bibinfo{author}{H.~Dekker}, 
  \bibinfo{title}{Classical and quantum mechanics of the damped harmonic oscillator}, 
  \bibinfo{journal}{Phys. Rep.} \bibinfo{volume}{{\bf 80}}, 
  \bibinfo{pages}{1} (\bibinfo{year}{1981}).
  
\bibitem{Exner1983}
\bibinfo{author}{P.~Exner}, 
  \bibinfo{title}{Complex-potential description of the damped harmonic oscillator}, 
  \bibinfo{journal}{J. Math. Phys.} \bibinfo{volume}{{\bf 24}}, 
  \bibinfo{pages}{1129} (\bibinfo{year}{1983}).
  
\bibitem{Baskoutas1993}
\bibinfo{author}{S.~Baskoutas}, \bibinfo{author}{A.~Jannussis}, and \bibinfo{author}{R.~Mignani}, 
  \bibinfo{title}{Time evolution of Caldirola-Kanai oscillators}, 
  \bibinfo{journal}{Il Nuovo Cimento B} \bibinfo{volume}{{\bf 108}}, 
  \bibinfo{pages}{953} (\bibinfo{year}{1993}).

\bibitem{Pedrosa1997}
\bibinfo{author}{I.~A. Pedrosa}, 
  \bibinfo{title}{Exact wave functions of a harmonic oscillator with time-dependent mass and frequency},
  \bibinfo{journal}{Phys. Rev. A} \bibinfo{volume}{{\bf 55}}, 
  \bibinfo{pages}{3219} (\bibinfo{year}{1997}).

\bibitem{Um2002}
\bibinfo{author}{C.-I. Um}, \bibinfo{author}{K.-H. Yeon}, and \bibinfo{author}{T.~F. George}, 
  \bibinfo{title}{The quantum damped harmonic oscillator}, 
  \bibinfo{journal}{Phys. Rep.} \bibinfo{volume}{{\bf 362}}, 
  \bibinfo{pages}{63} (\bibinfo{year}{2002}).

\bibitem{Razavy2006}
\bibinfo{author}{M.~Razavy}, 
  \bibinfo{title}{{\it Classical and Quantum Dissipative Systems}} 
  (\bibinfo{publisher}{Imperial College Press, London}, \bibinfo{year}{2006}).

\bibitem{Pepore2006}
\bibinfo{author}{S.~Pepore}, \bibinfo{author}{P.~Winotai}, \bibinfo{author}{T.~Osotchan}, and \bibinfo{author}{U.~Robkob},
  \bibinfo{title}{Path integral for a harmonic oscillator with time-dependent mass and frequency}, 
  \bibinfo{journal}{ScienceAsia} \bibinfo{volume}{{\bf 32}},
  \bibinfo{pages}{173} (\bibinfo{year}{2006}).

\bibitem{Cheng1988}
\bibinfo{author}{C.~M. Cheng} and \bibinfo{author}{P.~C.~W. Fung},
  \bibinfo{title}{The evolution operator technique in solving the Schr\"odinger equation, and its application to disentangling exponential operators and
  solving the problem of a mass-varying harmonic oscillator},
  \bibinfo{journal}{J. Phys. A} \bibinfo{volume}{{\bf 21}}, 
  \bibinfo{pages}{4115} (\bibinfo{year}{1988}).

\end{thebibliography}
\end{document}